\documentclass[trackchanges,twocolumn,twocolappendix]{aastex701} 

\usepackage[version=4]{mhchem}
\usepackage{graphicx}
\usepackage{subcaption}
\usepackage{soul}
\usepackage{tabularx}
\usepackage{multirow}%
\usepackage{booktabs} 
\usepackage{seqsplit}
\usepackage{threeparttable}
\usepackage{bm}
\usepackage{placeins}   
\usepackage{float}      
\allowdisplaybreaks     

\begin{document}

\title{Sub-Neptunes as Soot Factories: Deep Atmosphere Hydrocarbon Formation and Quenching as the Origin of Sub-Neptune Aerosol Trends}

\author[0000-0002-1551-2610]{Jeehyun Yang}
\affiliation{Department of Astronomy and Astrophysics, The University of Chicago, Chicago, IL 60637, USA}
\email[show]{jeehyuny@uchicago.edu}  

\author[0000-0002-1337-9051]{Eliza M.-R. Kempton}
\affiliation{Department of Astronomy and Astrophysics, The University of Chicago, Chicago, IL 60637, USA}
\email{ekempton@uchicago.edu}

\author[0000-0002-2454-768X]{Arjun B. Savel} 
\affiliation{Department of Astronomy, University of Maryland, College Park, 4296 Stadium Drive, College Park, MD 207842, USA}
\email{asavel@umd.edu}

\begin{abstract}

Recent population-level studies of sub-Neptune atmospheres have identified a tentative parabolic trend in transmission spectrum amplitude for planets with $T_{\rm eq}\approx500$$-$800 K.  While the trend has been commonly attributed to hydrocarbon aerosols, we lack a first-principles explanation of its underlying chemical mechanism. Previous work has focused on the role of methane photolysis and subsequent polymerization, but with limited reaction networks that truncated at \ce{C2}-species and couldn't reproduce the observed parabolic trend.  In this work, enabled by a computer-automated, rate-based chemical network generator, we construct the most comprehensive carbon reaction network for exoplanet atmospheres to date.  We explicitly model the formation of polycyclic aromatic hydrocarbons (PAHs), which are well established as soot precursors in combustion chemistry.  We calculate the chemical timescales of hydrocarbon species through an eigenvalue timescale method and model their quenched abundances across a range of C/O, metallicities, and $T_{\rm eq}$. In this framework, the deep atmosphere acts as a ``soot factory'' analogous to a combustion engine, transporting the ingredients for hydrocarbon aerosol formation to the JWST-observable region of the atmosphere,  where it may be further augmented by photochemistry.  We find that the predicted abundances of  PAHs peak near 600 K, and fall off toward higher and lower $T_{\rm eq}$, consistent with the observed muted-spectra regime suggested in observational studies by HST and JWST.  We also show that PAH abundances are expected to vary with C/O and metallicity, thus providing a natural explanation for observed diversity among planets with similar $T_{\rm eq}$.

\end{abstract}



\section{INTRODUCTION} \label{sec:introduction}

The formation of condensed-phase materials suspended in gas, often referred to as aerosols, has been studied across a wide range of environments, from candle flames, where nanoscale carbonaceous particles form, grow, and radiate through incandescence \citep{Faraday-1825-flame, faraday1861course}; to black smoke emitted by gasoline and diesel engine exhaust \citep{CLAGUE19991553}; to Earth’s water clouds and smog \citep{haagen1952chemistry}. Aerosols are also prevalent beyond Earth, appearing for example in the atmospheres of Jupiter \citep{west1981near}, Saturn \citep{TOMASKO19841}, and Titan \citep{tomasko1980preliminary}, along with other solar system planets and moons. More recently, strong circumstantial evidence has emerged for the presence of aerosols in exoplanet atmospheres \citep[e.g.,][]{Pont2013, Sing2016, gao2021aerosols}. In particular, observational studies of exoplanets have repeatedly been hindered by featureless or highly muted transmission spectra, with a likely cause being widespread aerosol particles that obscure atmospheric signals \citep{knutson2014featureless, kreidberg2014clouds}.

\begin{figure}[b!]
    \centering    
    \includegraphics[width=0.46\textwidth]{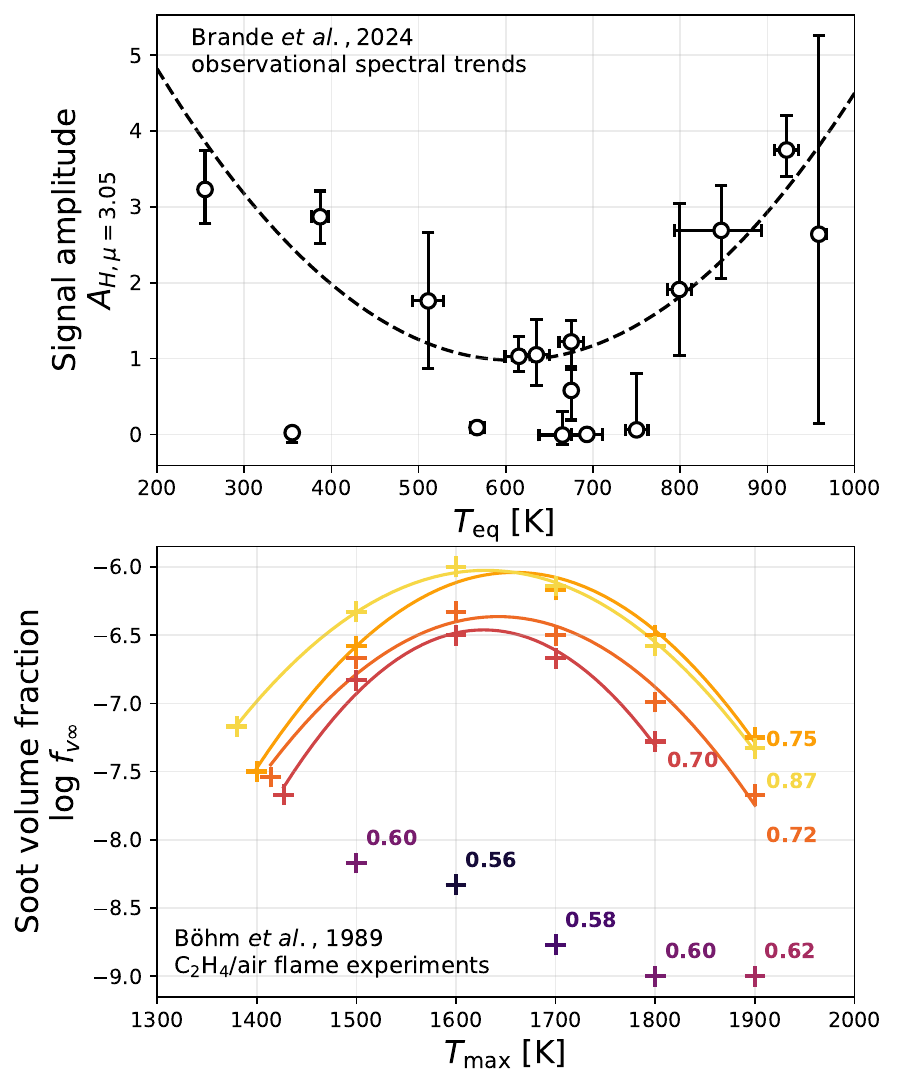}
     \caption{\footnotesize (Top)  Spectral amplitude at 1.4 $\mu$m $A_{H,\ \mu=3.05}$ vs.\ $T_{\rm eq}$, adapted from \citet{brande2024clouds} and  \citet{roy2025diversity}. The dashed line indicates the best-fit parabolic trend derived by \citet{brande2024clouds}. (Bottom)  Soot volume fraction log$(f_{\rm v\infty})$ vs.\ maximum flame temperature $T_{\rm max}$ adapted from \citet{BOHM1989403}. Numbers denote the corresponding experimental C/O conditions for the \ce{C2H4}/air flame experiments. Colored solid lines are shown for visual guidance to highlight the trend. Although derived from different scientific disciplines, the two panels suggest a possible correspondence between the processes driving both trends, and thus $T_{\rm eq}\sim$200$-$1000 K and $T_{\rm max}\sim$1300$-$2000K.}
    \label{fig:combustion_exoplanet_connection}
\end{figure}

Population-level studies of exoplanet atmospheres have also emerged as a powerful technique for constraining the presence and properties of aerosols. Recent population-level studies of sub-Neptune atmospheric observations show that muted and flat transmission spectra occur preferentially over a constrained range of planetary equilibrium temperatures ($T_{\rm eq}$) between 500 and 800 K and follow a tentative parabolic spectral trend with $T_{\rm eq}$ between 200 and 1000 K \citep[see the top panel of Figure~\ref{fig:combustion_exoplanet_connection};][]{crossfield2017trends, brande2024clouds, roy2025diversity}.

However, the existence and interpretation of this $T_{\rm eq}$-dependent trend remain uncertain. Recent studies \citep{Kahle-2025, roy2025diversity} have also identified sub-Neptune atmospheres that do not follow the proposed parabolic behavior (e.g., HD~86226~c and LP~791-18~c, which lie outside the 500–800 K range yet exhibit muted spectra), and suggest that reduced spectral feature amplitudes may instead arise from increased atmospheric mean molecular weight, without requiring aerosol opacity. These differing interpretations highlight that the physical and chemical drivers of spectral diversity in sub-Neptune atmospheres remain poorly constrained, and $T_{\rm eq}$ may not be the only parameter governing spectral amplitudes.  Of relevance, larger gas giant exoplanets also show similar $T_{\rm eq}$-dependent aerosol trends \citep{gao2020aerosol}, albeit with evidence for an upward shift in the transition temperature to clear atmosphere conditions for more massive planets \citep{Ashtari_2025}, further indicating the potential role of planet surface gravity or metallicity as a driver of aerosol properties.


While the sub-Neptune parabolic trend is only tentative due to small number statistics and likely diversity among the planet population, its confirmation and interpretation have received considerable attention.  It is generally believed in the exoplanet community that this temperature dependence reflects changes in the dominant aerosol composition. Aerosol microphysics models of giant exoplanet atmospheres suggest that hydrocarbon hazes dominate at $T_{\rm eq}\leq 950$ K, while silicates and other metal oxides become the primary aerosol constituents at $T_{\rm eq}\geq 950$ K \citep{gao2020aerosol}. This bifurcation arises from temperature-dependent differences in the aerosol production pathway.  High temperatures favor the condensation of silicates and metal oxide clouds \citep[e.g.,][]{Burrows1999, Visscher2010b, Mbarek2016}, whereas the basic ingredients for hydrocarbon haze formation (i.e., methane and its derivatives) become abundant at $T \lesssim 900$~K \citep{zahnle2009thermometric, Moses2013}.  From this perspective, any spectral trend observed over $T_{\rm eq} = 200$–1000 K appears to be linked to hydrocarbon aerosol formation.


Indeed, multiple previous studies have pointed to the formation of high-altitude hydrocarbon hazes driven by \ce{CH4} photolysis, analogous to the hazy atmosphere of Titan \citep{zahnle2009thermometric, kempton2011atmospheric, Morley_2013, Morley_2015, kawashima2018theoretical}. Among these, \citet{Morley_2013, Morley_2015} made use of 1D photochemical modeling to predict the dependence of haze ``precursor'' (\ce{C2H_{\rm x}} and \ce{HCN}) column densities on equilibrium temperature over the range $T_{\rm eq} \approx 450$-1400~K. Their models produce a peak in haze precursor abundance near 800 K, suggestive of the trend shown in the top panel of Figure~\ref{fig:combustion_exoplanet_connection}. However, several limitations of these existing calculations merit additional follow-up. For example, the chemical networks truncated at \ce{C2} hydrocarbons and relied on highly simplified parameterizations to predict the conversion of these small gas-phase molecules to aerosols. Furthermore, the existing models do not extend down to the lowest $T_{\rm{eq}}$ values of JWST-observed sub-Neptunes \citep[i.e., $\sim 260$~K for K2-18b;][]{Madhusudhan_2023}, nor do they explain the diversity in aerosol coverage seen among planets of similar  $T_{\rm eq}$ \citep[e.g., GJ~436~b vs.\ GJ 9827~d or LP~791-19~c vs.\ TOI-270~d;][]{knutson2014featureless, benneke2024jwst, Piaulet-Ghorayeb_2024, roy2025diversity}.

Recent studies have also explored the role of polycyclic aromatic hydrocarbons (PAHs) in exoplanet atmospheres from both thermochemical and observational perspectives. For example, \citet{dubey2023polycyclic} investigated PAH formation under thermochemical equilibrium in irradiated and non-irradiated giant-planet atmospheres. Subsequent work has examined the detectability of PAHs with JWST, both for individual targets \citep{grubel2025detectability} and across a broader range of atmospheric conditions \citep{arenales2025polycyclic}.

Interestingly, the combustion community has also studied a similar problem of ``soot'' (hydrocarbon aerosol) formation from gaseous fuel mixtures for several decades. To avoid ambiguity, we define ``soot'' as carbonaceous condensed-phase particles formed through high-temperature hydrocarbon chemistry, in which molecular growth proceeds via aromatic hydrocarbons (AHs) and polycyclic aromatic hydrocarbons (PAHs) \citep{wang2011formation, frenklach2020mechanism, MARTIN2022100956}. This is distinct from the organic hazes commonly discussed in planetary atmospheres, which are typically produced by photochemical processes in the upper atmosphere. In particular, laboratory analogs of such photochemical hazes, often referred to as tholins \citep{sagan1979tholins}, form through UV- or spark discharge-driven chemistry of simple molecules (e.g., \ce{CH4}, \ce{N2}). 

Although it is widely recognized that none of the existing soot models have full predictive power \citep{MARTIN2022100956}, largely due to uncertainties in gas-to-condensed phase inception kinetics and subsequent particle growth, this body of work has nevertheless produced extensive theoretical and experimental insight into the physical and chemical principles governing soot formation. One broad and robust consensus is that PAHs act as the key molecular precursors to soot particles \citep{wang2011formation, frenklach2020mechanism, MARTIN2022100956}. A central remaining question is which molecular species and reaction pathways control the transition from small gas molecules to PAHs and then to soot particles.

A well-established result addressing this question comes from premixed \ce{C2H4}/air flame experiments, which identified a threshold for soot formation and revealed a characteristic bell-shaped dependence of soot volume fraction on temperature, peaking near 1600 K \citep[see Figure~\ref{fig:combustion_exoplanet_connection}, bottom panel, modified from Fig.~3 of][]{BOHM1989403}. This behavior is naturally evocative of the tentative parabolic spectral trend observed in sub-Neptune atmospheres \citep[Figure~\ref{fig:combustion_exoplanet_connection}, top panel, modified from][]{crossfield2017trends, brande2024clouds, roy2025diversity}, suggesting that similar thermochemical principles may work in exoplanet atmospheres. Notably, another combustion experiment \citep{street1955carbon} also showed a strong dependence of soot formation on the C/O ratio, with a threshold onset near C/O$\geq$0.5, comparable to solar abundance values \citep{lodders2021relative}. This suggests that compositional parameters such as C/O may play an important role in regulating the hydrocarbon aerosol formation in exoplanet atmospheres (see the bottom panel of Figure~\ref{fig:combustion_exoplanet_connection}).

If one considers that the deep atmosphere temperatures for planets with $T_{\rm eq}$=400$-$800 K are expected to fall around 1500$-$1800 K based on Figure~\ref{fig:combustion_exoplanet_connection}, then these combustion and exoplanet phenomena can be meaningfully connected, and provide a physically grounded framework for identifying the thermochemical conditions under which PAH growth becomes efficient. This motivates our hypothesis that the molecular precursors of soot particles (which are PAHs) are produced thermochemically in the deep atmospheres of sub-Neptunes and subsequently transported upward by vertical mixing, where their abundances become quenched. Upon reaching the upper atmosphere, haze formation may be further enhanced by photochemical processing, making these aerosols observable. Accordingly, in this work, we test this hypothesis using a first-principles approach by explicitly modeling the reaction pathways from small molecules to large PAHs and applying these results in the context of exoplanet atmospheric observations.

\section{METHODS} \label{sec:method}
\subsection{Model Parameter Grids} \label{subsec:grids}
To explore the effects that govern PAH formation and quenching behavior in sub-Neptunes, we generated a large grid of model atmospheres, spanning a broadly relevant parameter space.
Our grid samples three key parameters: equilibrium temperature, $T_{\rm eq}$, ranging from 350 to 1200 K with 10 grid points; metallicity, [M/H]\footnote{in log$_{10}$ relative to solar}, ranging from 0.0 to 4.0 with 9 grid points; and carbon to oxygen ratio, C/O, with discrete values of 0.2, 0.55, 1.0, and 2.5 as shown in Table~A\ref{table:parameters}. The $T_{\rm eq}$ range is chosen to span the regime of sub-Neptune atmospheres observed by HST and JWST where hydrocarbon aerosols are expected to vary significantly. The lower bound of 350 K avoids regimes where water condensation becomes important, while the upper bound of 1200 K extends into conditions where silicate and metal oxide clouds may dominate. The metallicity range covers solar ([M/H]=0) to highly metal-enriched atmospheres, consistent with values inferred for sub-Neptune planets \citep[e.g., $\mathrm{[M/H]}\sim$3.5;][]{Ohno_2025}, and extends slightly beyond current observational constraints to allow broader exploration. The C/O range spans strongly oxygen-rich to highly carbon-rich compositions.

This resulted in a total of 360 model grid points ($10\times9\times4$). For the standard solar metallicity case [M/H] = 0, we adopted present-day solar abundances from Table 3 of \citet{lodders2021relative}. When varying C/O from the original solar value of 0.55 \citep{lodders2021relative}, we adjusted the relative abundances of carbon and oxygen while holding the total C+O abundance fixed. The resulting elemental abundances for each metallicity and C/O pairing are provided as a .CSV file in the supplementary materials (\texttt{abundance\_grids.csv}).

\subsection{The T--P profiles} \label{subsec:helios}
In our framework, the efficiency of PAH formation and quenching depends on the atmospheric thermal structure, as temperature primarily controls chemical reaction timescales and molecular growth processes. To explore the sensitivity of these processes to variations in thermal structure for realistic sub-Neptune conditions, we generated a grid of one-dimensional (1-D) radiative-convective equilibrium (RCE) models with the GPU-accelerated \texttt{HELIOS} code \citep{malik2017helios, malik2019self}.  We create a single 1-D temperature-pressure ($T$-$P$) profile at each $T_{\rm{eq}}$--[M/H]--C/O grid point listed in Table~A\ref{table:parameters}).  The radiative-convective calculation additionally requires the input of an intrinsic temperature ($T_{\rm int}$), which accounts for heat flux from the planet's deep interior, as detailed in Appendix~\ref{sec:tp_profile_appendix}.  
Additional details on the $T$-$P$ profiles are provided in Appendix~\ref{sec:tp_profile_appendix}. 

\subsection{Automatic Chemical Reaction Network Generation for PAH-chemistry up to \ce{C16}} \label{subsec:chemical_network}
We constructed a detailed chemical network describing complex PAH chemistry up to \ce{C16} using the Reaction Mechanism Generator \citep[\texttt{RMG};][]{Gao_2016, liu2021rmg, RMG-database}, an open-source Python package that automatically generates chemical networks. Previous exoplanet atmosphere models have generally limited hydrocarbon chemistry to small species (i.e., \ce{C2H}$_{\rm{x}}$), whereas our approach explicitly tracks the formation and growth of larger hydrocarbon molecules relevant to soot precursors. \texttt{RMG} has been widely applied to exoplanet atmosphere studies \citep{yang2024automated, Damiano_2024, yang2024chemical, benneke2024jwst, hu2025water, bello2025methane}, and more recently to modeling CO–\ce{CH4} chemistry in Jupiter \citep{yang2026jupiter}. The resulting chemical network consists of 700 species and 8258 reactions and is provided in the Supplementary Materials as a YAML file compatible with \texttt{Cantera} \citep{Goodwin_Cantera_An_Object-oriented_2024}, along with the corresponding species dictionary (\texttt{soot\_network\_v1.yaml} and \texttt{soot\_network\_v1.txt}). Additional details on the construction of the PAH chemical reaction network are provided in Appendix~\ref{sec:rmg_appendix}.

\subsection{Chemical Network Species Classification} \label{subsec:classification}
AH and PAH refer to classes of molecules rather than individual species. Because evaluating the quenching behavior of every species is computationally impractical, we assume that all AH species share the same quenching behavior (quench pressure) as a single grid-representative AH species, and likewise for PAH species. To implement this approach, we classified the 700 species generated by \texttt{RMG} (see  Section~\ref{subsec:chemical_network}) into three categories using the open-source cheminformatics toolkit \texttt{RDKit} \citep{greg_landrum_2025_rdkit}: 313 \texttt{no-ring} species, 157 \texttt{AH} species (containing one ring), and 230 \texttt{PAH} species (containing two or more rings). The full classification is provided in the supplementary materials as a CSV file (\texttt{species\_ring\_classification.csv}), accompanied by SVG files showing the molecular structures of the species constructed for this work.

We then identified representative species for the AH and PAH classes. For each of the 360 grid points considered in Section~\ref{subsec:grids}, we computed the thermochemical equilibrium abundances of all AH and PAH species along the $T$-$P$ profile computed in Section~\ref{subsec:helios}. Within each class, the species with the largest equilibrium mixing ratio at $P = 1$ mbar was selected as the representative species, and its quench pressure was assigned to all other species in the same class. For example, at C/O = 0.55, [M/H] = 2.5, and $T_{\rm eq} = 500$ K, benzene (\ce{C6H6}) and indane (\ce{C9H10}) are identified as the grid-representative AH and PAH species, respectively (see the middle panel of Figure~C\ref{fig:showcase}). Consequently, the AH and PAH group abundances inherit the quench pressures of benzene and indane, as shown in the right panel of Figure~C\ref{fig:showcase}.

\subsection{1D Quench Modeling Using a Characteristic Chemical Timescale Approach} \label{subsec:jacobian}
Vertical transport-induced quenching occurs when the chemical timescale, $\tau_{\mathrm{chem}}$, becomes comparable to the vertical mixing timescale, $\tau_{\mathrm{mix}}$, at a given pressure level \citep{prinn1977carbon}. Although the most rigorous way to model quenching in exoplanet atmospheres is to directly solve the full set of one-dimensional chemical kinetic-transport differential equations \citep{moses2011disequilibrium, venot2012chemical, Hu_2012, Rimmer_2016, Tsai_2017}, this approach becomes computationally impractical for large chemical networks. This is because the computational cost scales linearly with the number of reactions and quadratically with the number of species \citep{schwer2002upgrading}.

Given the large chemical network used in this work, consisting of 700 species and 8258 reactions as described in Section~\ref{subsec:chemical_network}, we instead adopt a characteristic chemical timescale approach. This method is well established in combustion science and provides an efficient way to approximate chemical timescales in complex systems. Several numerical formulations exist, including the eigenvalue timescale method (EVTS), inverse Jacobian timescale, system progress timescale, and progress variable timescale \citep{caudal-2013-chemicaltimescale, PRUFERT2014416}. Among these, EVTS generally provides higher accuracy \citep{Wartha10122021} and has been successfully applied to Jupiter’s atmosphere, reproducing a quenched CO abundance within a factor of two of full 1D kinetic–transport simulations \citep{yang2026jupiter}. Given the typical order-of-magnitude uncertainties in JWST-retrieved molecular abundances, the EVTS approximation is adequate for the present study and is adopted.

Using EVTS, described in detail in Appendix~\ref{sec:appendix_evts}, we computed the characteristic timescales of the species of interest along the $T$-$P$ profiles described in Section~\ref{subsec:helios}. We tested eddy diffusion coefficients of $K_{\rm zz} = 10^4$–$10^{8}$ cm$^2$/s and found that the resulting trends were not strongly sensitive to $K_{\rm zz}$. We therefore report only results using the nominal value of $10^6$ cm$^2$/s for sub-Neptunes \citep{Zhang&Showman2018} in this work. The quench level was identified as the highest pressure where $\tau_{\mathrm{chem}} \geq \tau_{\mathrm{mix}}$. If no crossing occurred, the species abundance was assumed to remain in thermochemical equilibrium. Figure~C\ref{fig:showcase} illustrates representative quenching behavior at $K_{\rm zz}=10^6$ [cm$^2$/s]  for one of our model atmosphere grid points  (Section~\ref{subsec:grids}) with C/O = 0.55, [M/H] = 2.5, and $T_{\rm eq} = 500$ K. Additional details of the EVTS implementation are provided in Appendix~\ref{sec:appendix_evts}.

\subsection{1D Photochemical Modeling of \ce{C2H2} Enhancement} \label{subsec:epacris}

The EVTS quench calculations described above account for disequilibrium via vertical mixing, but not photochemistry. To explore the potential enhancement of hydrocarbon formation due to photochemistry, we performed 1D photochemical kinetic–transport modeling for a subset of 10 of the model atmospheres from our total grid --- those with C/O = 0.55 and [M/H] = 2.5, while varying the $T_{\rm eq}$ from 350 to 1200 K. The simulations were carried out using \texttt{EPACRIS} \citep{yang2024automated}. We adopted the photochemical network from \citep{yang2024chemical} and assumed a constant $K_{\rm zz} = 10^6$ cm$^2$/s as a nominal value for sub-Neptunes \citep{Zhang&Showman2018}. The stellar flux at each $T_{\rm eq}$ was scaled according to the bolometric luminosity, using the M4.5V stellar spectrum of GJ 1214 from the MUSCLES survey\footnote{\url{https://cos.colorado.edu/~kevinf/muscles.html}} \citep{loyd_2016, France_2016}. We ran models both with and without photochemical reactions to assess their impact on the \ce{C2H2} mixing ratio at $P = 1$ mbar, which represents the endpoint of hydrocarbon chemistry in the photochemical network. We define the photochemical enhancement factor ($f_{\rm photo}$) for \ce{C2H2} as the ratio of its mixing ratio with photochemistry to that without photochemistry at the same $T_{\rm eq}$.

\begin{figure*}[htb!]
    \includegraphics[width=\textwidth]{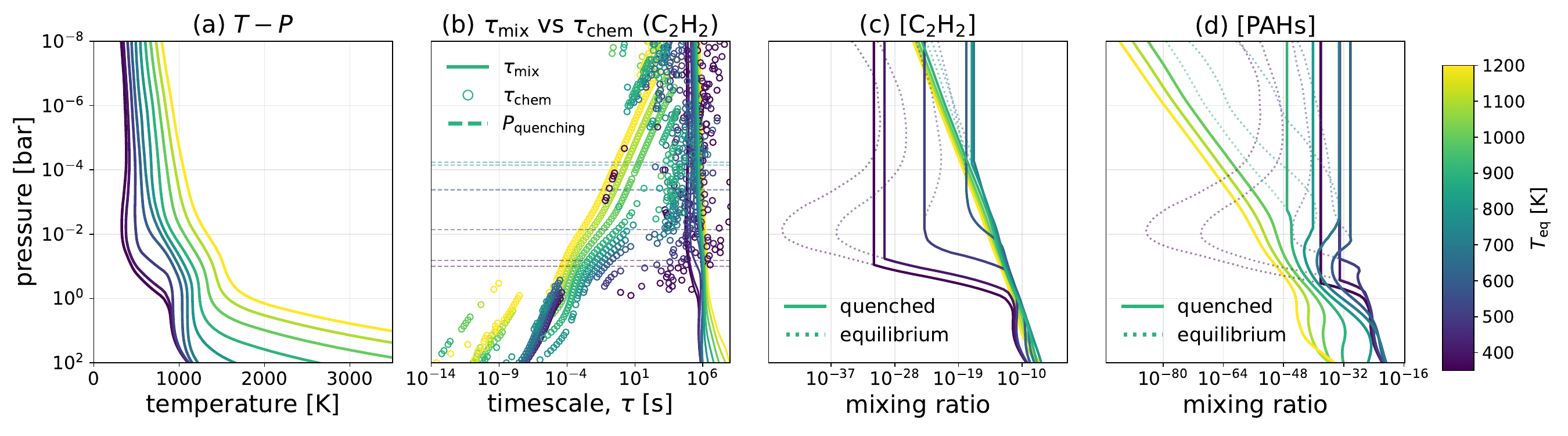}
     \caption{\footnotesize Detailed quenching behavior as a function of $T_{\rm eq}$=350–1200 K at fixed C/O=0.55 and [M/H]=2.5. Colors indicate $T_{\rm eq}$. (a) Temperature–pressure ($T$–$P$) profiles for each $T_{\rm eq}$. (b) Timescale–pressure ($\tau$–$P$) profiles. Open circles denote the chemical timescale of \ce{C2H2} evaluated along the corresponding $T$–$P$ profile in panel (a), while solid lines indicate the vertical mixing timescale. Horizontal dashed lines mark the quench pressure. No quenching occurs for \ce{C2H2} at $T_{\rm eq}\geq$800 K. (c) Vertical mixing ratio profiles of \ce{C2H2}. Solid lines show quenched abundances, while dotted lines show thermochemical equilibrium profiles. (d) Same as (c), but for PAHs.}
    \label{fig:c2h2_pah_teq}
\end{figure*}

\section{RESULTS AND DISCUSSION} \label{sec:results_discussion}
\subsection{Similar Temperature Regimes in Combustion Systems and Deep Sub-Neptune Atmospheres}\label{thermal_structures} 
Figure~\ref{fig:c2h2_pah_teq} shows an example subset of 10 models with fixed C/O = 0.55 and [M/H] = 2.5, varying \mbox{$T_{\rm eq} = 350$–1200 K}, drawn from the full grid of 360 models explored in this study. As a reminder, for each grid point, we first compute the temperature–pressure ($T$–$P$) profile, shown in panel (a), following Section~\ref{subsec:helios}. Using these profiles, we then calculate the chemical timescales of the target species and their corresponding quench points, shown in panel (b), following Section~\ref{subsec:jacobian} and Appendix~\ref{sec:appendix_evts}). This gives us vertical mixing ratio profiles for each species, for example, \ce{C2H2} and PAHs in panels (c) and (d), respectively. The resulting formation and quenching behavior across all 360 grid conditions is summarized in Figure~\ref{fig:2D_soot_grid_Kzz_1e6}.

A key result taken from Figure~\ref{fig:c2h2_pah_teq}a is that the modeled thermal structures with $T_{\rm eq}$=600$–$800 K reach deep atmospheric temperatures of approximately 1200–1800 K. This temperature range closely overlaps with the temperatures characteristic of soot-producing combustion experiments, as shown in Figure~\ref{fig:combustion_exoplanet_connection}. Herein lies the direct connection between the temperature regime of the combustion experiment results ($T_{\rm max}$ in Figure~\ref{fig:combustion_exoplanet_connection}) and the JWST and HST observations of sub-Neptune atmospheres.  There is a near-one-to-one correspondence between the $T_{\rm max}$ values of peak soot production from the flame experiments and the deep-atmosphere temperatures of sub-Neptunes with $T_{\rm eq} $$\approx$ 600~K, where transmission spectra are observed to be the most muted.  The similarity in thermochemical regimes supports the applicability of combustion-based intuition for interpreting soot precursor formation in deep sub-Neptune atmospheres.

\subsection{Soot Precursors in Sub-Neptune Atmospheres}\label{subsec:precursor_thermochemistry_quenching}
\begin{figure*}[htb!]
    \includegraphics[width=1\textwidth]{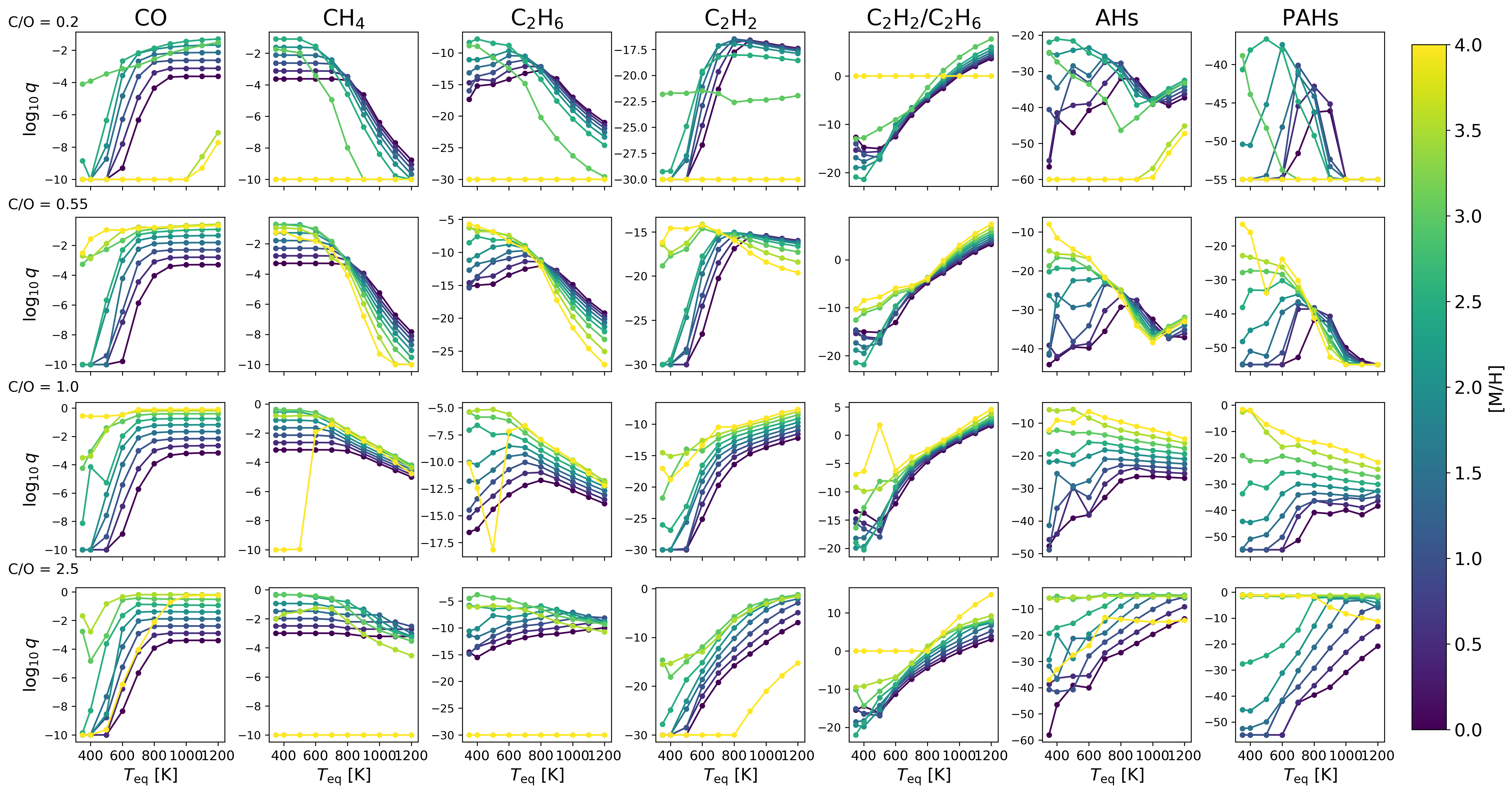}
     \caption{\footnotesize Overall formation and quenching behavior for multiple carbon-bearing chemical species across 360 atmospheric conditions (Section~\ref{subsec:grids}). In each panel, the x-axis shows the $T_{\rm eq}$, and the y-axis shows the logarithm of the quenched volume mixing ratio at $P$=1 mbar (log$_{10} q$). Columns correspond to CO, \ce{CH4}, \ce{C2H6}, \ce{C2H2}, \ce{C2H2}/\ce{C2H6}, AHs, and PAHs. Rows correspond to C/O=0.2, 0.55, 1.0, and 2.0 from top to bottom. Colors indicate metallicity, [M/H], spanning from 0 (solar) to 4.}
    \label{fig:2D_soot_grid_Kzz_1e6}
\end{figure*}

Figure~\ref{fig:2D_soot_grid_Kzz_1e6} presents the combined effect of thermochemical formation and quenching behavior of several key species simulated at $P = 1$ mbar, representative of the JWST-observable region of exoplanet atmospheres \citep{rustamkulov2023early}, across our full grid of 360 models (Section~\ref{subsec:grids}). In the following subsections, we interpret these results from a thermochemistry and quenching perspective in a stepwise manner. We also briefly discuss the potential role of photochemistry on soot precursors in sub-Neptune atmospheres.

\subsubsection{A Thermochemistry Perspective}\label{subsubsec:thermochemistry_perspective}
From left to right in Figure~\ref{fig:2D_soot_grid_Kzz_1e6}, we observe a general trend with increasing $T_{\rm eq}$ toward progressive molecular growth of hydrocarbon species, from the simplest form, \ce{CH4}, to PAHs. At lower $T_{\rm eq} \leq 800$ K, \ce{CH4} dominates over CO due to its thermochemical stability in the deep atmosphere, followed by quenching, a process that has been well studied by \citet{moses2011disequilibrium}.

For \ce{C2} species, the saturated molecule \ce{C2H6} increases with $T_{\rm eq}$ up to approximately 800 K and then decreases at higher $T_{\rm eq} \geq 800$ K. This behavior arises because \ce{C2H6} formation primarily proceeds through the recombination of \ce{CH3} radicals stemming from \ce{CH4}. At higher $T_{\rm eq}$, \ce{CH4} becomes depleted \citep{moses2011disequilibrium}, suppressing \ce{CH3} production and, consequently, \ce{C2H6} formation. In addition, $T_{\rm eq}$ above 800 K is sufficiently high to promote H-abstraction and $\beta$-scission, which thermally crack saturated hydrocarbons (alkanes such as \ce{C2H6}) into unsaturated species (alkenes and alkynes such as \ce{C2H2}) via hydrogen loss. As a result, unsaturated \ce{C2} species such as \ce{C2H2} dominate over \ce{C2H6} at higher $T_{\rm eq}$, as shown in the fifth column of Figure~\ref{fig:2D_soot_grid_Kzz_1e6}. More broadly, the shift from saturated to unsaturated \ce{C2} species highlights the role of multiple carbon-carbon bonds in PAH formation. Saturated hydrocarbons with single \ce{C-C} bonds are relatively inert toward ring growth, whereas unsaturated species with \ce{C=C} or \ce{C#C} bonds, particularly \ce{C2H2}, are chemically active intermediates that efficiently drive aromatic growth through hydrogen-abstraction/acetylene-addition (HACA) reactions \citep{frenklach1983soot}.

Beyond \ce{C2} species, the chemistry becomes more complex and is no longer a simple monotonic function of $T_{\rm eq}$. Instead, hydrocarbon growth results from a balance of thermochemical processes driven by either enthalpy or entropy, depending on temperature, pressure, and initial composition. One good example that shows this behavior is the experimental study of \ce{C11H22} (Jet-A fuel) pyrolysis at $T = 1300$ K and $P = 10$ bar by \citet{WANG2018502}. \citet{WANG2018502} showed that early decomposition steps up to alkynes are entropy-driven as large molecules fragment into smaller species, whereas the first aromatic ring formation is enthalpy-driven. However, subsequent polycyclic aromatic hydrocarbon growth again becomes entropy-driven, as illustrated in Figure 6 of \citet{WANG2018502}.

This complex thermochemical behavior implies that soot precursor formation requires temperatures high enough to sustain molecular growth, while excessively high temperatures then fragment key intermediates needed for PAH formation. This competition is what drives the bell-shaped dependence of soot formation on temperature observed in combustion experiments (bottom panel of Figure~\ref{fig:showcase}). In our model grid, the dependence of AH and PAH abundances on $T_{\rm eq}$ is non-monotonic only in specific compositional regimes, particularly for C/O $\leq 0.55$ and moderate metallicities ($\mathrm{[M/H]} \lesssim 3.5$), while other regimes are dominated by metallicity-driven monotonic trends (Figure~\ref{fig:2D_soot_grid_Kzz_1e6}).

The C/O ratio and metallicity, [M/H], also play important roles by controlling the abundance of carbon and oxygen atoms (in combustion, C/O is often described in terms of the fuel-to-air ratio). Together, these parameters determine the oxidative state of the system \citep{BOHM1989403, GLASSMAN1989295}. Increasing C/O generally promotes soot formation by enhancing the availability of carbon relative to oxygen, favoring the production of unsaturated hydrocarbon species. In contrast, metallicity exerts a non-monotonic influence, promoting soot formation up to a threshold beyond which the chemistry becomes increasingly oxidizing.

At elevated metallicity, the absolute carbon abundance increases, which initially favors PAH formation. In the extreme case of [M/H] = 4 and C/O = 2.5, as shown in Figure~\ref{fig:2D_soot_grid_Kzz_1e6}, the composition becomes strongly hydrogen-limited, with carbon significantly enhanced relative to hydrogen ($\sim$58\% C, $\sim$10\% H, and $\sim$23\% O by number). Under these conditions, saturated hydrocarbons such as \ce{CH4} and \ce{C2H6} are suppressed due to limited hydrogen availability, and excess carbon is instead incorporated into more unsaturated species, mainly PAHs, some aromatic hydrocarbons, and some \ce{C2H2}, enhancing the production of soot precursors.

However, as metallicity continues to increase, the concurrent rise in oxygen abundance introduces a competing oxidizing effect. This enhanced oxygen abundance efficiently terminates carbon-chain growth pathways \citep{GLASSMAN1989295}, suppressing PAH formation. This effect dominates in oxygen-rich regimes (e.g., C/O = 0.2) at high metallicity ([M/H] $\gtrsim$ 3.5), where carbon fully oxidizes into \ce{CO2} (also known as complete combustion), as indicated by the yellow and light green lines in Figure~\ref{fig:2D_soot_grid_Kzz_1e6}. As a result, hydrocarbon species spanning from \ce{CH4} to PAHs are strongly suppressed.

\subsubsection{A Quenching Perspective}\label{subsubsec:quenching_perspective}
While atmospheric temperature is a key driver of planetary atmospheric chemistry, vertical transport induced by convection or diffusion can drive atmospheres away from thermochemical equilibrium and into disequilibrium chemistry \citep{moses2011disequilibrium}. Thus, it is important to account for the effects of vertical mixing in exoplanet atmospheric modeling. Figure~\ref{fig:c2h2_pah_teq}b--d illustrates quenching behavior under different $T_{\rm eq}$ scenarios for exoplanets with fixed C/O = 0.55 and [M/H] = 2.5.

One notable trend in Figures~\ref{fig:c2h2_pah_teq}b is that as $T_{\rm eq}$ increases, chemical quenching occurs at higher altitudes (i.e., lower pressures). At $T_{\rm eq} \gtrsim 700$ K, neither \ce{C2H2} nor PAHs achieve quenching over the entire modeled atmosphere (see Figure~\ref{fig:c2h2_pah_teq}c and d). This behavior arises because at higher $T_{\rm eq}$, thermochemical reaction rates become much faster than vertical mixing. In other words, the chemical timescale ($\tau_{\rm chem}$) becomes much shorter than the mixing timescale ($\tau_{\rm mix}$), as shown in Figure~\ref{fig:c2h2_pah_teq}b. As a result, the atmospheric composition closely follows thermochemical equilibrium throughout the model domain. This regime is typical of hot and ultra-hot Jupiters, as demonstrated in previous modeling studies of HD~209458~b \citep{moses2011disequilibrium}. We note that our model assumes a uniform $K_{\rm zz}$, which is unlikely to be realistic for actual exoplanet atmospheres. However, given that $K_{\rm zz}$ remains poorly constrained in planetary science, this simplified treatment nonetheless captures the essential qualitative quenching behavior of chemical species and thus provides useful insight.

Turning to Figure~\ref{fig:c2h2_pah_teq}d, the effects of quenching lead to substantially enhanced vertical mixing ratios of PAHs at 1 mbar, the pressure level typically probed by JWST, compared to their thermochemical equilibrium abundances. As a result, PAHs are expected to be significantly more abundant at lower $T_{\rm eq}$, with a pronounced peak at $T_{\rm eq} = 600$ K. In contrast, \ce{C2H2} exhibits a more monotonic dependence on $T_{\rm eq}$, with its mixing ratio increasing with temperature and reaching a plateau at $T_{\rm eq} \approx 700$ K. This highlights that quenching plays an important role in setting PAH abundances under the grid conditions explored in this study. We remind the reader that variations in $K_{\rm zz}$ (see Appendix~\ref{sec:appendix_evts}) have a weaker effect on the bell-shaped trend than changes in $T_{\rm eq}$.

\subsubsection{A Photochemistry Perspective}\label{subsubsec:photochemistry_perspective}
Although photochemistry is known to play a major role in shaping hydrocarbon abundances and initiating aerosol formation, as demonstrated by observations of Titan’s atmosphere \citep{yung1984photochemistry} and laboratory experiments simulating exoplanet conditions \citep{He_2018, fleury2019photochemistry, yu2021haze}, most full chemical kinetic models restrict haze precursors to \ce{C2H}$_{\rm x}$ species and HCN due to incomplete understanding of photochemical hydrocarbon growth pathways and computational limitations \citep{zahnle2009thermometric, Tsai_2021, Huang_2024}. Consistent with these constraints, our photochemical network includes carbon-bearing species only up to \ce{C2}. While photochemistry is also likely to significantly enhance the production of PAHs, these processes remain poorly understood and are beyond the scope of the present study. Accordingly, in the following section, we mainly consider PAH formation through thermochemical pathways. Photochemical effects on \ce{C2H2} are explicitly evaluated for the case with C/O = 0.55, and [M/H] = 2.5, and the same photochemical enhancement is assumed to apply across the remainder of the parameter grid.

\begin{figure}[htb!]
    \centering    
    \includegraphics[width=0.45\textwidth]{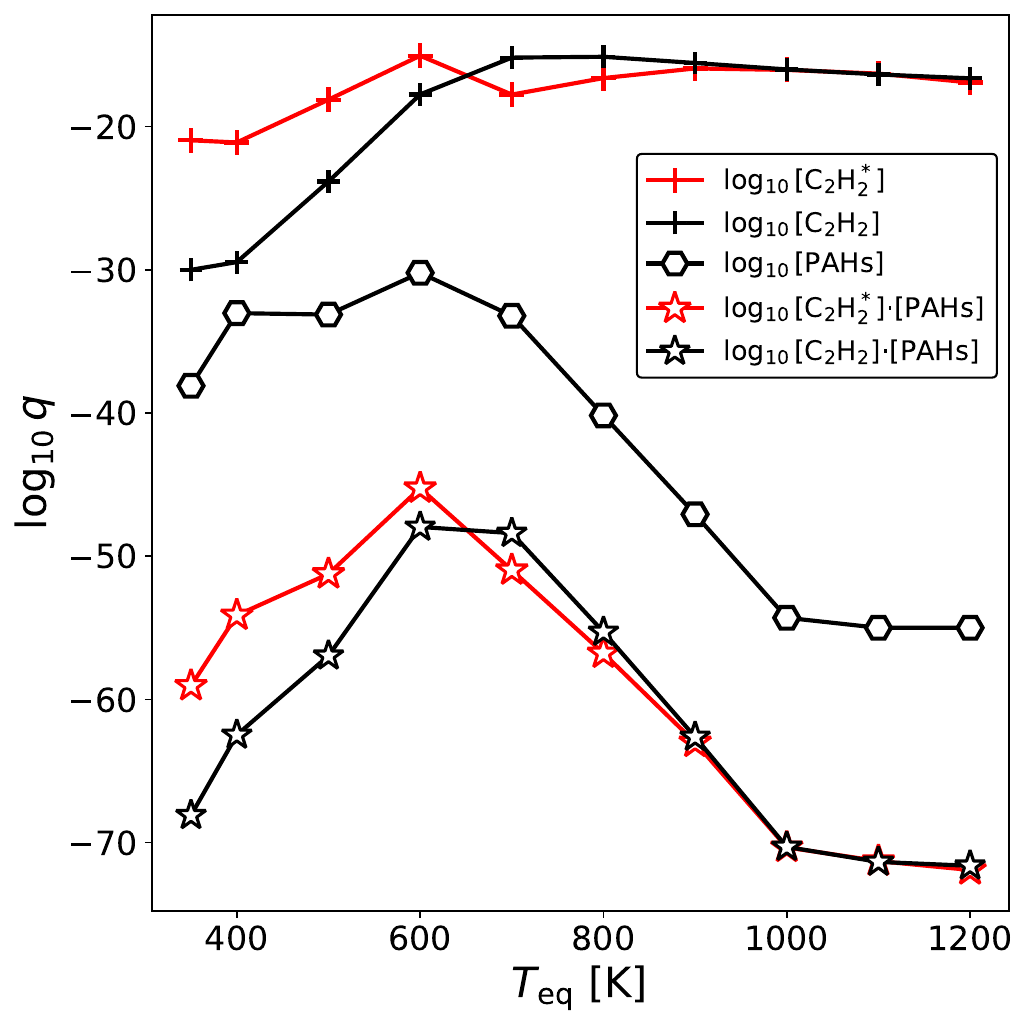}
     \caption{\footnotesize Photochemically enhanced \ce{C2H2} formation for the C/O=0.55 and [M/H]=2.5 case. Crosses, hexagons, and stars denote the logarithms of the \ce{C2H2} mixing ratio, PAH mixing ratio, and $[\ce{C2H2}]\cdot[\mathrm{PAHs}]$, respectively, all evaluated at $P$=1 mbar. Black symbols show abundances obtained from quench chemistry alone. Red symbols marked with $^{*}$ show the same quench-chemistry abundances scaled by the photochemical enhancement factor $f_{\rm photo}$, derived from separate photochemical modeling described in Section~\ref{subsec:epacris}. These values therefore represent photochemically enhanced quench abundances rather than direct outputs of the photochemical simulations.}
    \label{fig:photochemistry_enhancement}
\end{figure}

As shown in Figure~\ref{fig:photochemistry_enhancement}, photochemistry enhances \ce{C2H2} abundances (red crosses) by up to 10 orders of magnitude at $T_{\rm eq} \leq 600$ K compared to the thermochemistry-only case. The black crosses represent \ce{C2H2} abundances obtained from quench chemistry alone, while the red crosses show these same quench-chemistry abundances scaled by the photochemical enhancement factor, $f_{\rm photo}$, derived from separate photochemical modeling described in Section~\ref{subsec:epacris}. This approach isolates the effect of photochemistry on top of the quenched \ce{C2H2} abundances. The photochemical enhancement becomes progressively weaker at higher $T_{\rm eq}$, where thermochemistry dominates \citep{yang2023metastableCO}. This enhancement at lower $T_{\rm eq}$ is therefore expected to substantially promote soot formation through surface growth and HACA pathways, relative to thermochemistry alone.

\subsection{Sooting Propensity of Sub-Neptune Atmospheres}\label{subsec:sooting_propensity}

As discussed in the Section~\ref{sec:introduction}, there is a broad consensus within the combustion community that alkynes such as \ce{C2H2} and polycyclic aromatic hydrocarbons (PAHs) are positively correlated with the formation of soot particles, often described as ``\textit{Sooting Propensity}'', based on combustion experiments \citep{MARTIN2022100956}. However, none of the existing soot models are predictive, as aerosol formation involves a transition from gas-phase to condensed-phase growth that is not well understood by either experimental kinetics or first-principles theory \citep{MARTIN2022100956}. In contrast to gas-phase reactions, which can be treated using well-established first-principles approaches such as transition state theory \citep{Eyring-1935} and the Rice–Ramsperger–Kassel–Marcus theory \citep{Marcus-1951, Marcus_1952}, the kinetics of condensed-phase particle formation remain poorly understood. Nevertheless, insights from combustion chemistry can still provide a useful framework for interpreting trends in exoplanet atmospheric composition.

Motivated by this perspective, we introduce a heuristic metric for the sooting propensity of sub-Neptune atmospheres, denoted by $\eta$, intended as a relative indicator rather than a predictive measure of aerosol formation. We define $\eta$ as
\begin{equation}
\eta = \frac{([\ce{C2H2}]\cdot[\mathrm{PAHs}])}{([\ce{C2H2}]\cdot[\mathrm{PAHs}])_{\mathrm{ref}}},
\end{equation}
where the subscript $\mathrm{ref}$ denotes a reference value associated with an exoplanet known to exhibit muted spectral features due to aerosols. We adopt the product [\ce{C2H2}]$\cdot$[PAHs] rather than either abundance alone, because these species play complementary roles in soot formation. \ce{C2H2} provides the dominant feedstock for surface growth through HACA \citep{frenklach1983soot}, while PAHs serve as aromatic nuclei or growth surfaces in combustion systems \citep{MARTIN2022100956}. As shown in Figure~\ref{fig:photochemistry_enhancement}, \ce{C2H2} is often more abundant. However, HACA-driven growth alone becomes kinetically inefficient beyond naphthalene (2-ring PAH), as demonstrated by both theoretical and experimental studies \citep{kislov2013formation, yang2021}. Efficient formation of larger PAHs and soot particles therefore requires the presence of \textit{existing} PAHs \citep{wang2011formation, frenklach2020mechanism, MARTIN2022100956}. Their product therefore provides a heuristic measure of the simultaneous availability of both growth feedstock and growth surfaces, and is more illustrative of sooting propensity than either abundance considered in isolation. We adopt GJ~1214~b as this reference case, as it represents the canonical example of a sub-Neptune with a flat transmission spectrum attributed to aerosols, supported by multiple observations from ground-based telescopes \citep{bean2010ground, Bean_2011}, HST \citep{Berta_2012, kreidberg2014clouds}, and JWST \citep{kempton2023reflective, Schlawin_2024, Ohno_2025}. Given GJ~1214~b’s $T_{\rm eq} = 567 \pm 8$ K \citep{Mahajan_2024} and [M/H] $\sim 3.5$ \citep{kempton2023reflective, Ohno_2025}, we adopt the modeled value of [\ce{C2H2}]$\cdot$[PAHs] at C/O = 0.55, [M/H] = 3.5, and $T_{\rm eq} = 600$ K as the reference value.

As an illustrative example, Figure~\ref{fig:photochemistry_enhancement} shows results for C/O = 0.55 and [M/H] = 2.5, where black crosses denote [\ce{C2H2}], open black circles denote [PAHs], and black stars indicate their product, $[\ce{C2H2}]\cdot[\mathrm{PAHs}]$, at the corresponding $T_{\rm eq}$. To account for photochemical enhancement of \ce{C2H2}, quantities that include the photochemistry effect are denoted with an asterisk, for example $[\ce{C2H2}^*] = [\ce{C2H2}] \times  f_{\rm{photo}}$. Dividing these values by the reference value yields $\eta$, which should be interpreted as a relative measure of sooting propensity normalized to a GJ~1214~b-like atmosphere under the modeled conditions defined by a given C/O, [M/H], and $T_{\rm eq}$. 

\begin{figure}[htb!]
    \includegraphics[width=0.47\textwidth]{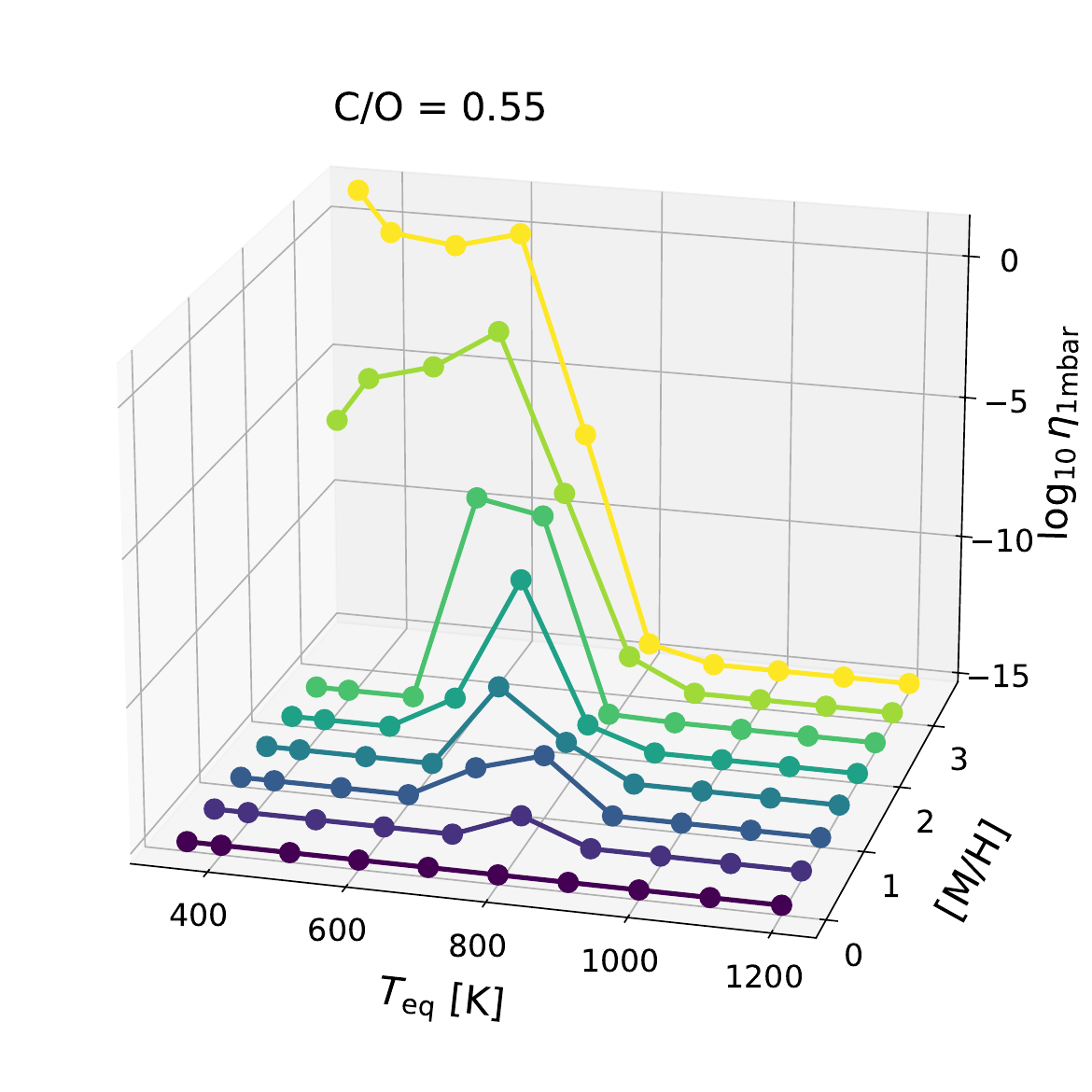}
     \caption{\footnotesize 3D plot of $T_{\rm eq}-$[M/H]$-\log_{10}\eta_{1 \rm mbar}$ at fixed C/O=0.55 \citep[i.e., solar C/O;][]{lodders2021relative}. The definition of the sooting propensity, $\eta$, is described in Section~\ref{subsec:sooting_propensity}.}
    \label{fig:3d_sooting_propensity}
\end{figure}

The resulting sooting propensity over a range of [M/H] and $T_{\rm eq}$ values at fixed C/O = 0.55 is shown in Figure~\ref{fig:3d_sooting_propensity}. Two notable features emerge. First, the maximum value of $\eta$ is 12.54, while the next highest value of $\eta$=1 corresponding to a sooting propensity comparable to that of GJ~1214~b, which has [M/H] = 3.5 and $T_{\rm eq}$ = 600 K. Assuming a solar C/O ratio, this places GJ~1214~b at the extreme of the explored parameter space in terms of its propensity for soot formation, providing a natural explanation for its role as a benchmark case for haze-dominated atmospheres.

Second, a bell-shaped dependence on $T_{\rm eq}$ emerges at metallicities of [M/H] $\geq 0.5$, becoming apparent by [M/H] $\sim 2.5$ and peaking near $T_{\rm eq}$ = 600 K. At higher metallicities ([M/H] $\geq 3$), the sooting propensity increases rapidly even at $T_{\rm eq} \leq 500$ K, and the bell-shaped trend disappears, giving way to a monotonic decline with increasing $T_{\rm eq}$.

In this framework, higher values of $\eta$ indicate atmospheric conditions that are more favorable for hydrocarbon aerosol production and therefore more likely to yield featureless transmission spectra at the pressure levels probed by JWST ($P \sim 1$ mbar). Atmospheres with $\eta \sim 1$ are thus expected to exhibit flat transmission spectra similar to that of GJ~1214~b. In the next section, we apply this sooting propensity metric to existing exoplanet observations to examine its observational implications.

\subsection{Implication for Atmospheric Observations}\label{subsec:Implications}

\begin{figure*}[htb!]
    \includegraphics[width=1\textwidth]{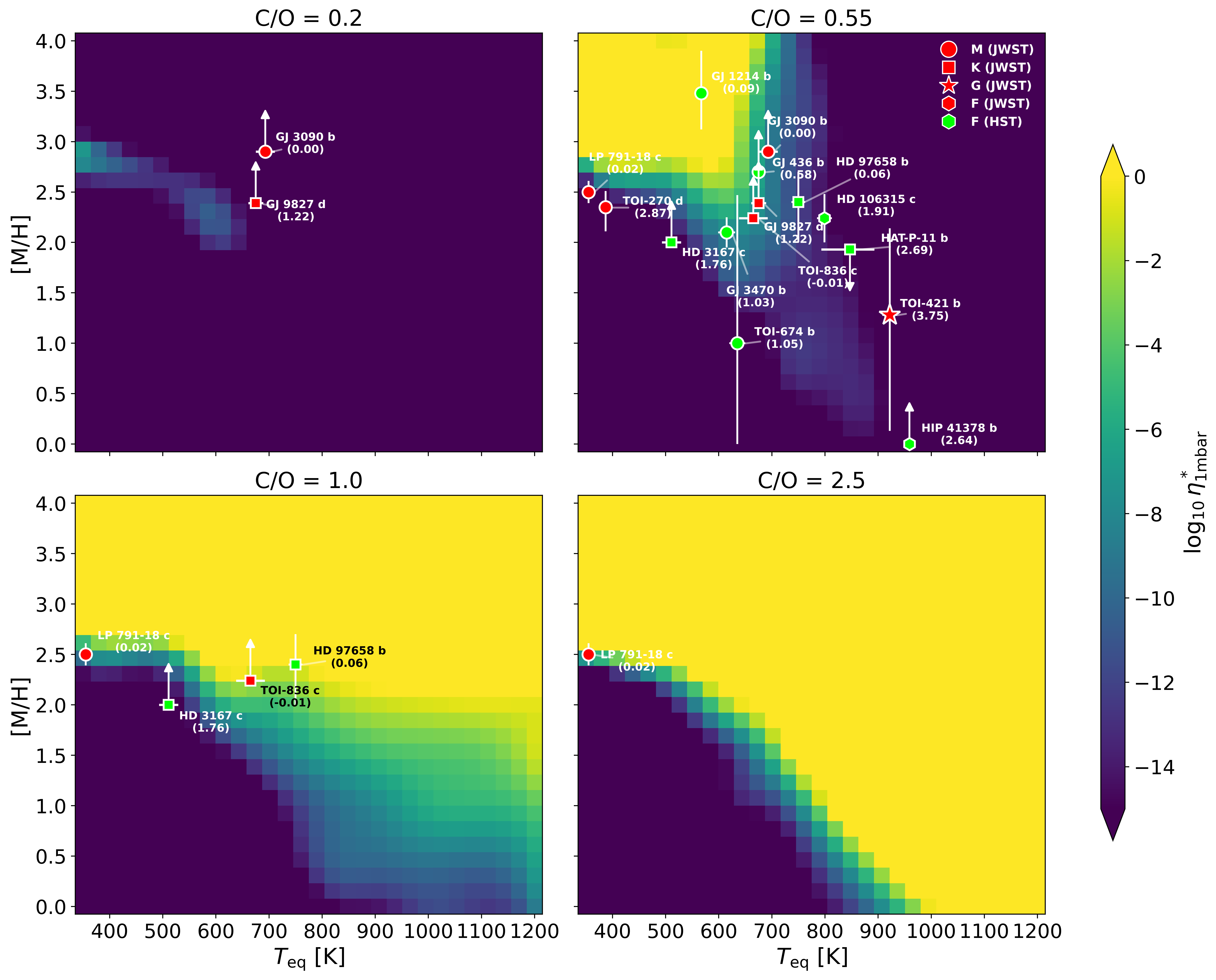}
     \caption{\footnotesize 2D color map as a function of $T_{\rm eq}$ and [M/H], where the color denotes the logarithm of the photochemistry effect-included sooting propensity at $P$=1 mbar, $\eta^*_{\rm 1mbar}$, as defined in Section~\ref{subsec:sooting_propensity}. Each panel corresponds to a different C/O, spanning 0.2 to 2.5, as indicated. Markers with error bars indicate the observations for which [M/H] constraints have been reported in the literature. All planets are assumed to have a solar C/O ratio of 0.55 by default and are plotted in the upper right-hand panel. However, for planets with indications from the literature that their atmospheric C/O may deviate from the solar value (LP~791-18~c, HD~3167~c, GJ~3090~b, GJ~9827~d, TOI-836~c, and HD~97658~b), we additionally plot their locations in the corresponding non-solar C/O panels to illustrate the sensitivity of their interpretation to the assumed C/O ratio. Numbers in parentheses denote the signal amplitude of the 1.4~$\mu$m \ce{H2O} or \ce{CH4} absorption band in the corresponding observations, expressed in units of atmospheric scale heights and assuming a mean molecular weight of 3.05 adopted from \citet{brande2024clouds, roy2025diversity}. Color denotes the observatory from which the signal amplitude was derived, with JWST shown in red and HST shown in lime. Marker shapes denote host star spectral type, with circles for M type, squares for K type, stars for G type, and hexagons for F type. References and adopted atmospheric parameters for each plotted planet are provided in Table~D\ref{table:planet_data} in Appendix~\ref{sec:parametric_data_table}.}
    \label{fig:2D-colormap-JWST}
\end{figure*}

Figure~\ref{fig:2D-colormap-JWST} shows a 2D color map of the photochemistry-enhanced sooting propensity at $P=1$ mbar, $\eta^*_{\rm 1mbar}$, as a function of $T_{\rm eq}$, [M/H], and C/O, together with previous HST and JWST observations of various sub-Neptune exoplanets. A notable feature of Figure~\ref{fig:2D-colormap-JWST} is the presence of a clear band of enhanced sooting propensity spanning $T_{\rm eq} \sim 550$–800 K at [M/H] $\geq 1$ for C/O = 0.55. This band extends to lower $T_{\rm eq}$ values at higher metallicities ([M/H] $\geq 2.5$), consistent with the trends discussed in the previous section and shown in Figure~\ref{fig:3d_sooting_propensity}, of which this figure represents a two-dimensional projection.  Below, we discuss on a planet-by-planet basis how existing observations of individual sub-Neptunes fit into this picture.

\subsubsection{GJ~1214~b}\label{subsubsec:gj1214b}
Figure~\ref{fig:2D-colormap-JWST} provides a convenient visualization of the regions in $T_{\rm eq}$–[M/H] space that are most favorable for haze formation driven by deep atmospheric PAH production corresponding to high sooting propensity\footnote{We define soot formation regimes based on $\eta^*_{1\rm mbar}$ as follows: high ($\geq 10^{-5}$), moderate ($10^{-10}$-$10^{-5}$), and low ($\leq 10^{-10}$).}. In this context, GJ~1214~b \citep{kreidberg2014clouds, kempton2023reflective, Schlawin_2024, Ohno_2025} falls within our predicted regions of highest $\eta$, consistent with GJ~1214~b's observed near-zero signal amplitude and muted transmission spectrum. Notably, our predicted \ce{CO2} abundance of $\sim$0.54 and mean molecular weight of $\sim$31 [g/mol] are consistent with recent JWST-based studies \citep{Schlawin_2024, Ohno_2025}, which infer a \ce{CO2}-rich, high–mean molecular weight atmosphere with a \ce{CO2} abundance between 0.4 and 0.5 \citep[see Figure 8 of ][]{Ohno_2025}. This agreement provides an independent consistency check on our framework and supports the interpretation that thick hydrocarbon hazes may be responsible for suppressing spectral features in GJ~1214~b's atmosphere. In fact, GJ 1214~b is the only planet with a (near-) flat transmission spectrum observed to-date at sufficient observational precision to break the well-known mean molecular weight -- aerosol degeneracy \citep{Benneke2013}, thus confirming the presence of thick aerosols in its atmosphere.

In general, sub-Neptunes occupying similar regions of parameter space, with $T_{\rm eq}\sim 600$~K, solar or super-solar C/O, and [M/H] $\gtrsim 3$ are expected to exhibit the strongest sooting propensity in their deep atmospheres, making them more likely to display flat transmission spectra similar to that of GJ~1214~b \citep{bean2010ground, kreidberg2014clouds, kempton2023reflective, Schlawin_2024, Ohno_2025}. For GJ~1214~b, the atmospheric C/O ratio remains unconstrained due to its exceptionally flat transmission spectrum. Within our framework, however, reproducing the observed level of spectral suppression via hydrocarbon aerosols would require at least solar C/O.

\subsubsection{GJ~3090~b}\label{subsubsec:gj3090b}
For GJ~3090~b with its near-zero signal amplitude, however, the mean molecular weight$-$aerosol degeneracy cannot be fully broken at current observational precision \citep{Ahrer_2025}. It is therefore unclear whether the planet's muted spectrum arises from aerosol coverage, high mean molecular weight, or a combination of these effects. Within our framework, assuming C/O=0.55, GJ~3090~b has $\eta^*_{\rm 1mbar}\geq10^{-5}$, suggesting that some level of hydrocarbon haze formation is plausible and could contribute to the observed spectral muting. However, current constraints on the atmospheric C/O ratio \citep[C/O$<$0.54;][]{Ahrer_2025} also allow for substantially lower values. In such a scenario, despite a high formally inferred metallicity of [M/H] $\geq$ 2.89, the C/O = 0.2 panel of Figure~\ref{fig:2D-colormap-JWST} shows $\eta^*_{\rm 1mbar} \leq 10^{-15}$, indicating that soot formation would be strongly suppressed. In this case, the low signal amplitude is more likely explained by a high mean molecular weight atmosphere, potentially dominated by species such as \ce{H2O}, consistent with an extremely volatile-rich composition (e.g., a water world).  Future observations with JWST at higher precision could be useful for breaking the degeneracy among these scenarios.

\subsubsection{TOI-836~c}\label{subsubsec:toi836c}
A similar situation applies to TOI-836~c, which exhibits a signal amplitude of $-0.01\ H_{\mu=3.05}$ (consistent with zero, within error bars). Although atmospheric metallicities below 175$\times$ solar are ruled out for this planet in the aerosol-free case, the relative contributions of haze formation versus a high mean molecular weight atmosphere to its featureless transmission spectrum remain unclear \citep{wallack2024jwst}. Unlike GJ~3090~b, the C/O ratio of TOI-836~c is entirely unconstrained by current observations \citep{wallack2024jwst}. In the higher-C/O regime, as illustrated in the C/O = 1 panel of Figure~\ref{fig:2D-colormap-JWST}, TOI-836~c readily falls into a region with $\eta^*_{1\rm mbar} \sim 1$. Within the sooting propensity framework, this implies that substantial hydrocarbon haze formation is plausible for TOI~836~c if either [M/H] or C/O is sufficiently high, consistent with the microphysical modeling results reported by \citet{wallack2024jwst}.

\subsubsection{HD~97658~b}\label{subsubsec:hd97658b}
In contrast, HD~97658~b presents a discrepancy in the C/O=0.55 plane. Despite its low signal amplitude of $0.06\ H_{\mu=3.05}$ \citep{Guo_2020}, our framework predicts a relatively low sooting propensity when evaluated at C/O = 0.55. However, retrieval results from \citet{Guo_2020} suggest that HD~97658~b may have C/O $\geq 0.8$. When the planet is instead placed on the C/O=1 panel as shown in Figure~\ref{fig:2D-colormap-JWST}, HD~97658~b falls within a region of high sooting propensity, comparable to that of GJ~1214~b, bringing the model predictions into agreement with observations. In this regime, the atmosphere is consistent with a ``soot world'' interpretation, as suggested by \citet{Bergin_2023} and \citet{Li_2026}. More generally, for C/O $\geq 1$, our framework predicts that atmospheres with [M/H] $\geq 2.5$ exhibit $\eta^*_{1\rm mbar} \sim 1$ largely independent of $T_{\rm eq}$, implying persistently hazy conditions comparable to or more opaque than GJ~1214~b, and muted transmission spectra. This behavior is chemically intuitive, as C/O $\geq 1$ corresponds to a fuel-rich regime, analogous to a high fuel-to-air equivalence ratio ($\phi \geq3$) in combustion chemistry, where efficient soot formation during engine combustion is well known \citep{graham1975formation, frenklach1983soot, GLASSMAN1989295, BOHM1989403, tree2007soot}. This highlights another key result of our study, in which soot formation in carbon-rich atmospheres (i.e., C/O$\geq$1.0) becomes largely insensitive to $T_{\rm eq}$ and is instead primarily controlled by composition ([M/H]$\geq$2.5). 

\subsubsection{LP~791-18~c}\label{subsubsec:lp79118c}
A recent JWST NIRSpec/PRISM observation of LP~791-18~c with the signal amplitude of $\sim$0.02 and a high metallicity of 246–415$\times$ solar presents an unexpected result \citep{roy2025diversity}. Despite the planet’s low equilibrium temperature ($T_{\rm eq}=355$ K), a regime generally thought to be free of hazes based on observations of planets such as K2-18~b \citep{Madhusudhan_2023, hu2025water} and TOI-270~d \citep{benneke2024jwst}, the atmosphere was found to be highly hazy. This observation challenges the temperature-driven parabolic trends in atmospheric clarity reported by \citet{brande2024clouds} and points to significant diversity in the atmospheric properties of temperate sub-Neptunes. 

Interestingly, the sooting propensity framework provides a natural context for interpreting this result. In the C/O = 0.55 panel of Figure~\ref{fig:2D-colormap-JWST}, LP~791-18~c lies near the metallicity boundary between low and high $\eta^*_{1\rm mbar}$ regimes. This suggests that [M/H] plays a critical role in determining whether $\eta^*_{1\rm mbar}\gtrsim 1$, and therefore whether the atmosphere becomes hazy. This sensitivity is especially important given that the planet’s C/O ratio is poorly constrained observationally \citep{roy2025diversity}. As noted by \citet{roy2025diversity}, \ce{CO2} absorption is not detected, while \ce{CH4} is clearly present, yielding a \ce{CO2}-to-\ce{CH4} ratio below 0.07 at the $2\sigma$ level. Such a ratio suggests a low \ce{H2O} abundance in the deep envelope and is consistent with a high C/O ratio, indicative of relatively dry formation interior to the water ice line \citep{yang2024chemical}.

If LP~791-18~c is instead considered for C/O$\geq1$ as shown in C/O=1 and 2.5 panels of Figure~\ref{fig:2D-colormap-JWST}, it resides more firmly within a region of high sooting propensity. Given the recently inferred high C/O of $\sim 2.9$ for Jupiter \citep{yang2026jupiter}, an elevated C/O for LP~791-18~c is plausible. In this regime, efficient hydrocarbon haze production is expected, providing a straightforward explanation for the observed haziness despite the planet’s low $T_{\rm eq}$. At the same time, LP~791-18~c is observed to be less haze-dominated than GJ~1214~b, as evidenced by the presence of spectral features such as \ce{CH4} \citep{roy2025diversity}. This behavior is consistent with our model predictions, which indicate that LP~791-18~c has an $\eta$ value at least five orders of magnitude lower than that of GJ~1214~b in the C/O = 1 regime, naturally producing a hazy atmosphere that nonetheless retains detectable molecular features. More broadly, this example demonstrates that $T_{\rm eq}$ alone is not a sufficient predictor of atmospheric haziness. The combined influence of metallicity and C/O ratio must be considered when interpreting and forecasting the atmospheric properties of temperate sub-Neptunes in forthcoming JWST studies.

\subsubsection{GJ~436~b}\label{subsubsec:gj436b}
The nature of GJ~436~b’s atmosphere remains uncertain, as its JWST panchromatic thermal emission spectrum allows for both a cloudy, metal-enriched atmosphere with [M/H]$\geq$2.5 and a cloud-free atmosphere with lower metallicity \citep{mukherjee2025jwst}. However, the featureless transmission spectrum observed with HST \citep{knutson2014featureless} and VLT \citep{grasser2024peering} is more consistent with the presence of aerosols in a high metallicity atmosphere. Within our soot propensity framework, assuming a solar C/O ratio and [M/H]$\geq$2.5 places GJ~436~b in a high sooting propensity regime, providing a plausible explanation for its muted spectral features via soot formation. As with GJ~1214~b, HD~97658~b, and LP~791-18~c, this interpretation disfavors substantially sub-solar C/O ratios, which would move GJ~436~b into a low-sooting regime inconsistent with its featureless transmission spectrum. Additional JWST transmission spectroscopy will be essential for resolving the degeneracy between clear and hazy atmospheric scenarios by tightening constraints on both C/O and [M/H].

\subsubsection{GJ~9827~d}\label{subsubsec:gj9827d}
The origin of the low signal amplitude of GJ~9827~d (1.22) remains ambiguous, as it may result from hydrocarbon haze or clouds, or a high mean molecular weight atmosphere \citep{Piaulet-Ghorayeb_2024}. Assuming a solar C/O=0.55, the retrieved [M/H] and $T_{\rm eq}$ of GJ~9827~d place it in a moderate sooting propensity regime, offering a plausible explanation for its suppressed spectral features via soot formation as shown in the C/O=0.55 panel of Figure~\ref{fig:2D-colormap-JWST}.

Alternatively, the detection of strong \ce{H2O} absorption together with the non-detection of \ce{CH4} and evidence for helium escape from JWST NIRISS/SOSS observation may favor a steam world atmosphere with sub-solar C/O \citep{Piaulet-Ghorayeb_2024}. In this case, soot formation would be strongly suppressed across all metallicities, with $\eta^*_{1\rm mbar}\geq10^{-15}$ as shown in the \mbox{C/O = 0.2} panel of Figure~\ref{fig:2D-colormap-JWST}. The low signal amplitude of GJ~9827~d is therefore more naturally explained by a high mean molecular weight atmosphere rather than by soot formation. However, it should be noted that this inference is strongly driven by the current NIRISS observations, which have limited wavelength coverage \citep{Piaulet-Ghorayeb_2024} and therefore leave room for alternative explanations involving aerosols. Further atmospheric characterization of GJ~9827~d will be necessary to distinguish between these scenarios. In particular, upcoming transit observations with JWST NIRSpec/G395H \citep[GO 4098;][]{benneke2023exploring} targeting carbon and sulfur-bearing species would provide key constraints on C/O and the existence of hazes \citep{Piaulet-Ghorayeb_2024}.

\subsubsection{GJ~3470~b}\label{subsubsec:gj3470b}
GJ~3470~b appears to be less cloudy than similar mass-range exoplanets orbiting bright stars, such as GJ~1214~b and GJ~436~b \citep{Beatty_2024}. Although its signal amplitude of 1.03 is low \citep{brande2024clouds}, it is not as suppressed as those of GJ~1214~b \citep[0.09;][]{brande2024clouds} and GJ~436~b \citep[0.58;][]{brande2024clouds}, raising an interesting question about the physical origin of the differing levels of haziness among these three planets, which has remained unresolved \citep{Beatty_2024}. Recent JWST NIRCam observations of GJ~3470~b suggest a [M/H]$\sim$2 and C/O$\sim$0.55 \citep{Beatty_2024}. As shown in the C/O = 0.55 panel of Figure~\ref{fig:2D-colormap-JWST}, these properties place GJ~3470~b in a moderate sooting propensity regime ($\eta^*_{1\rm mbar}\gtrsim 10^{-7}$). Its $\eta^*_{1\rm mbar}$ is comparable to that of GJ~9827~d, whose signal intensity is 1.22, but at least an order of magnitude lower than that of GJ~436~b and approximately six orders of magnitude lower than that of GJ~1214~b. This hierarchy naturally explains the observed trend in haziness among these similar mass planets, with GJ~1214~b exhibiting the highest haze opacity, followed by GJ~436~b and then GJ~3470~b.

\subsubsection{TOI-674~b}\label{subsubsec:toi674b}
TOI-674~b is one of the relatively few known exo-Neptunes located within the Neptune desert, the observed paucity of Neptune-sized exoplanets at short orbital periods \citep{Brande_2022}. HST observations of TOI-674~b revealed a detectable \ce{H2O} vapor feature with a signal amplitude of 1.05 \citep{brande2024clouds}, along with tentative evidence for the presence of aerosols \citep{Brande_2022}. However, owing to the limited wavelength coverage of these observations, the inferred atmospheric metallicity remains highly degenerate, spanning a broad range from 1 to 300$\times$solar, with no meaningful constraints on the C/O ratio \citep{Brande_2022}. When interpreted within the sooting propensity framework, TOI-674~b is consistent with a moderate sooting regime if its metallicity is comparable to that of GJ~3470~b, which is suggested by their similar signal amplitudes near unity. As shown in the C/O=0.5 panel of Figure~\ref{fig:2D-colormap-JWST}, this scenario provides a plausible explanation for the observed spectral features. Upcoming JWST observations that combine both transmission and emission spectroscopy will be important for constraining the atmospheric metallicity and assessing the presence of aerosols in TOI-674~b’s atmosphere \citep[GO 9095;][]{piaulet2025combining}.

\subsubsection{Other Exoplanets}\label{subsubsec:otherexoplanets}
The sooting propensity framework provides a consistent explanation for all exoplanets with clear atmospheres observed to date. As shown in the C/O = 0.55 panel of Figure~\ref{fig:2D-colormap-JWST}, TOI-270~d, HD~3167~c, HD~106315~c, HAT-P-11~b, TOI-421~b, and HIP~41378~b all lie in regions of low sooting propensity and exhibit signal amplitudes greater than $1.76\ H_{\mu=3.05}$. For HD~3167~c and HD~106315~c, if the metallicity exceeds 200$\times$ solar ([M/H] $\geq$ 2.3), these planets begin to enter regions of moderate sooting propensity, which may explain why their observed signal amplitudes do not exceed 2.00. In contrast, TOI-270~d, HAT-P-11~b, TOI-421~b, and HIP~41378~b reside in the unambiguously clear regime where $\eta^*_{1\rm mbar} \leq 10^{-15}$, naturally explaining their aerosol-free atmospheres as revealed by JWST. Notably, even accounting for uncertainties in its metallicity, TOI-421~b remains in the soot-clear region across the full plausible [M/H] range, consistent with TOI-421~b's highest signal intensity of 3.75 observed by JWST so far \citep{davenport2025toi}. 

Among these planets, with the exception of TOI-270~d, the atmospheric C/O ratios are not yet well constrained. Within our sooting propensity framework, however, C/O ratios exceeding unity would shift HD~106315~c, HAT-P-11~b, TOI-421~b, and HIP~41378~b into higher sooting propensity regimes, which would be inconsistent with their observed large signal amplitudes. This therefore suggests that these planets are unlikely to possess substantially super-solar C/O ratios, providing an indirect constraint on their atmospheric composition within this framework. As additional JWST and HST observations expand the population of well-characterized exoplanet atmospheres, the sooting propensity framework can be systematically tested and refined. This positions the framework as both a robust interpretive tool for existing data and a predictive tool for selecting optimal targets for future atmospheric characterization.

\section{Conclusions} \label{sec:conclusions}

In this work, we combined modeling approaches from planetary science and combustion chemistry to connect sub-Neptune deep atmospheric thermochemistry with upper atmospheric observations by JWST and HST. By combining radiative transfer, chemical timescale analysis, and rate-based automated chemical network generation, we provide a first-principles explanation that can account for enhanced aerosol formation in $T_{\rm eq}$=500$-$800 K range, potentially contributing to the muted-spectra regime suggested by some populational studies \citep[e.g.,][]{crossfield2017trends, brande2024clouds} and identify its plausible origin in deep-atmospheric PAH formation.

In this framework, the deep atmosphere acts as a ``soot factory'' analogous to a gasoline or diesel combustion engine, in which PAHs and soot are efficiently produced and transported upward, amplified by photochemistry, and ultimately manifesting as an observed non-monotonic spectral trend. This framework consistently explains existing observations and establishes both a basis for interpretation and a predictive model for guiding future observations. For example, our results indicate that, at a solar C/O ratio, GJ~1214~b–like conditions with $T_{\rm eq}\sim600$ K and [M/H]$\sim3.5$ maximize soot propensity, making such planets a benchmark case for hydrocarbon haze-rich atmospheres. More generally, we find that in carbon-rich atmospheres (C/O$\geq$1), soot formation becomes largely insensitive to $T_{\rm eq}$ and is instead primarily controlled by composition, with high-metallicity conditions ([M/H] $\geq$ 2.5) leading to persistently hazy atmospheres, comparable to or more opaque than GJ~1214~b, across a wide range of $T_{\rm eq}$.

Our work suggests that hydrocarbon haze production is regulated not only by $T_{\rm eq}$, but also by atmospheric metallicity and C/O. The C/O ratio plays a role analogous to the fuel-to-air ratio in combustion chemistry, with higher C/O favoring incomplete combustion and the formation of unsaturated hydrocarbons and PAHs, thereby promoting haze formation. Increasing metallicity likewise enhances haze production by increasing the total carbon abundance. However, at very high metallicity ([M/H]$\geq$4), the accompanying increase in oxygen shifts the redox balance toward oxidized species such as \ce{CO2}, suppressing hydrocarbon formation in a regime analogous to complete combustion. These compositional axes add important nuance to the interpretation of sub-Neptune transmission spectra and naturally explain the differing degree of aerosol coverage inferred for planets with similar $T_{\rm eq}$, such as the hazy LP~791-18~c compared with the clear TOI-270~d (or K2-18~b).

Beyond providing a unified interpretive framework, our sooting propensity metric also yields testable predictions. Among moderate-to-high metallicity sub-Neptunes, the presence or absence of hydrocarbon hazes may serve as an indirect indicator of atmospheric C/O, distinguishing carbon-rich ``soot worlds'' from ``water worlds'' with sub-solar C/O. Because atmospheric C/O is closely linked to formation location and accretion history \citep{oberg2011effects}, haze availability may therefore provide a practical observational constraint on planet formation pathways \citep{Bergin_2023}. These predictions can be tested with JWST transmission spectroscopy by jointly constraining aerosol opacity and molecular abundances.

Future work should extend the framework developed in this Letter by incorporating a more detailed treatment of photochemical effects on PAH formation and by conducting a comprehensive kinetic investigation of the transition from gas phase to condensed phase, commonly referred to as ``particle inception''. The models produced in this work can also be coupled in the future to radiative transfer and aerosol microphysics calculations, in order to directly predict their observational consequences --- e.g., to determine what values of the sooting propensity $\eta$ are expected to produce clear, muted, or flat transmission spectra.
\begin{acknowledgments}
The authors thank Dr.\ Caroline Piaulet-Ghorayeb and Dr.\ Michael Radica for helpful discussions on the atmospheric observations of GJ~9827~d and GJ~3090~b.  E.M.-R.K.\ acknowledges support from the NASA Exoplanets Research Program under grant 80NSSC24K0157
\end{acknowledgments}

\begin{contribution}
J.Y. conceptualized and led the project. J.Y. and E.M.-R.K. jointly designed the study. J.Y. led the writing of the manuscript. J.Y. constructed the chemical network (\texttt{RMG}), computed the characteristic chemical timescales (\texttt{Cantera} and \texttt{scipy}), and conducted the 1D photochemical kinetic-transport modeling (\texttt{EPACRIS}). A.B.S. conducted 1D radiative-convective equilibrium calculation (\texttt{HELIOS}). All authors contributed to the writing and revision of the manuscript.
\end{contribution}

\software{ \texttt{HELIOS} \citep{malik2017helios}, \texttt{RMG} \citep{Gao_2016, liu2021rmg, RMG-database}, \texttt{RDKit} \citep{greg_landrum_2025_rdkit}, \texttt{Cantera} \citep{Goodwin_Cantera_An_Object-oriented_2024}, \texttt{scipy} \citep{2020SciPy-NMeth}, \texttt{EPACRIS} \citep{yang2024automated}}

\appendix
\setcounter{figure}{0}
\setcounter{table}{0}
\section{Details of the $T$-$P$ Profile Calculation}
\label{sec:tp_profile_appendix}

When simulating $T$-$P$ profiles with \texttt{HELIOS}, we systematically vary the planets’ $T_{\rm int}$, as soot formation in the deep atmosphere is expected to be sensitive to temperature. For hot Jupiters, it has been established how the intrinsic temperature varies as a function of $T_{\rm eq}$ \citep{thorngren_2019}. However, it is less clear how the energy deposition mechanisms that control $T_{\rm int}$ scale in the sub-Neptune regime. We therefore consider two regimes: a fixed $T_{\rm int} = 40$ K for $T_{\rm eq} \leq 600$ K, and a  $T_{\rm int}$ that depends on $T_{\rm eq}$ for $T_{\rm eq} \geq 700$ K following \cite{thorngren_2019}:
\begin{equation}\label{eq:thorngren}
    T_{\rm int} = 0.39\ T_{\rm eq}\ \rm exp\left(-\frac{(\log_{10}(\textit{F}/10^9) - 0.14)^2 }{1.095}\right),
\end{equation}
where $F$ is the flux incident on the planet as a function of $T_{\rm eq}$, expressed in erg/cm$^2$/s. The calculated $T_{\rm int}$ values are listed in Table~A\ref{table:parameters}. If the \cite{thorngren_2019} expression were extended to calculate $T_{\rm int}$ for the cooler planets, the temperature would fall below $40$~K (less than 1~K for the coolest model in our grid). Our constant value of 40~K approximates the  \cite{thorngren_2019} calculation for a GJ 1214-like planet.

Finally, we vary the grid based on metallicity and C/O. These quantities control the opacities available in the gas phase in the atmosphere, which in turn set local heating and cooling rates that set the thermal structure. We use the \texttt{FastChem} code \citep{stock2018fastchem, stock2022fastchem} to calculate the abundances of gas species as a function of metallicity, C/O, temperature, and pressure. We include opacity from a range of species that are expected to manifest in sub-Neptune atmospheres: \ce{CO2}, \ce{H2O}, \ce{CO}, \ce{Na}, \ce{K}, \ce{C2H2}, \ce{CH4}, \ce{NH3}, \ce{H2S}, \ce{OH} and \ce{HCN}. In addition, our calculations consider collision-induced absorption from \ce{H2-He} and \ce{H2-H2} pairs and scattering from \ce{He} and \ce{H2}. \texttt{HELIOS} then interpolates within this grid to set the local mean molecular weight and volume mixing ratios as a function within each modeled cell.

We note that radiative transfer and disequilibrium chemistry are not fully coupled in a self-consistent iterative framework. This approximation is justified in the deep atmosphere, where compositions follow thermochemical equilibrium. At higher altitudes, disequilibrium processes such as quenching can modify abundances and, in principle, affect the thermal structure. Previous work has shown that such effects are minimal at pressures greater than $\sim$0.1 bar, with temperature differences typically less than 100 K in the upper atmosphere ($P \leq 0.1$ bar) \citep{mukherjee2025effects}.

We note, however, that PAHs and potentially soot formed in the deep atmosphere, particularly under conditions of elevated C/O$\geq$1.0 and higher metallicity ([M/H] $\geq$3.5), may introduce additional opacity and radiative feedback in the upper atmosphere, which could lead to larger temperature deviations. The magnitude of this effect remains highly uncertain, as the optical properties of these species are poorly constrained and require further laboratory measurements. Fully self-consistent coupling between radiative transfer and disequilibrium chemistry remains an important direction for future work, but is not yet standard in the field and is beyond the scope of the current study.

\begin{deluxetable}{lc}[htb!]
\renewcommand{\tablename}{Table A}
\tablecaption{Parameter Ranges for Atmospheric Models\label{table:parameters}}
\tablehead{
\colhead{Parameter} & \colhead{Values}
}
\startdata
$T_\mathrm{eq}$ [K] & 350, 400, 500, 600, 700,\\
& 800, 900, 1000, 1100, 1200 \\
C/O & 0.2, 0.55 (solar), 1, 2.5 \\
$[$M/H$]$ & 0, 0.5, 1, 1.5, 2, 2.5, 3, 3.5, 4 \\
$T_\mathrm{int}$ [K]\tablenotemark{b} & 40\tablenotemark{a}, 45, 89, 150, 220, 300, 380 \\
\enddata
\tablenotetext{a}{$T_{\rm eq}\leq 600$ K}
\tablenotetext{b}{For $T_{\rm eq}\geq 700$ K, we vary $T_\mathrm{int}$ as a function of $T_\mathrm{eq}$ per Eq.~\ref{eq:thorngren}.}
\end{deluxetable}

\begin{figure*}[t!]
    \renewcommand{\figurename}{Figure C}
    \includegraphics[width=1\textwidth]{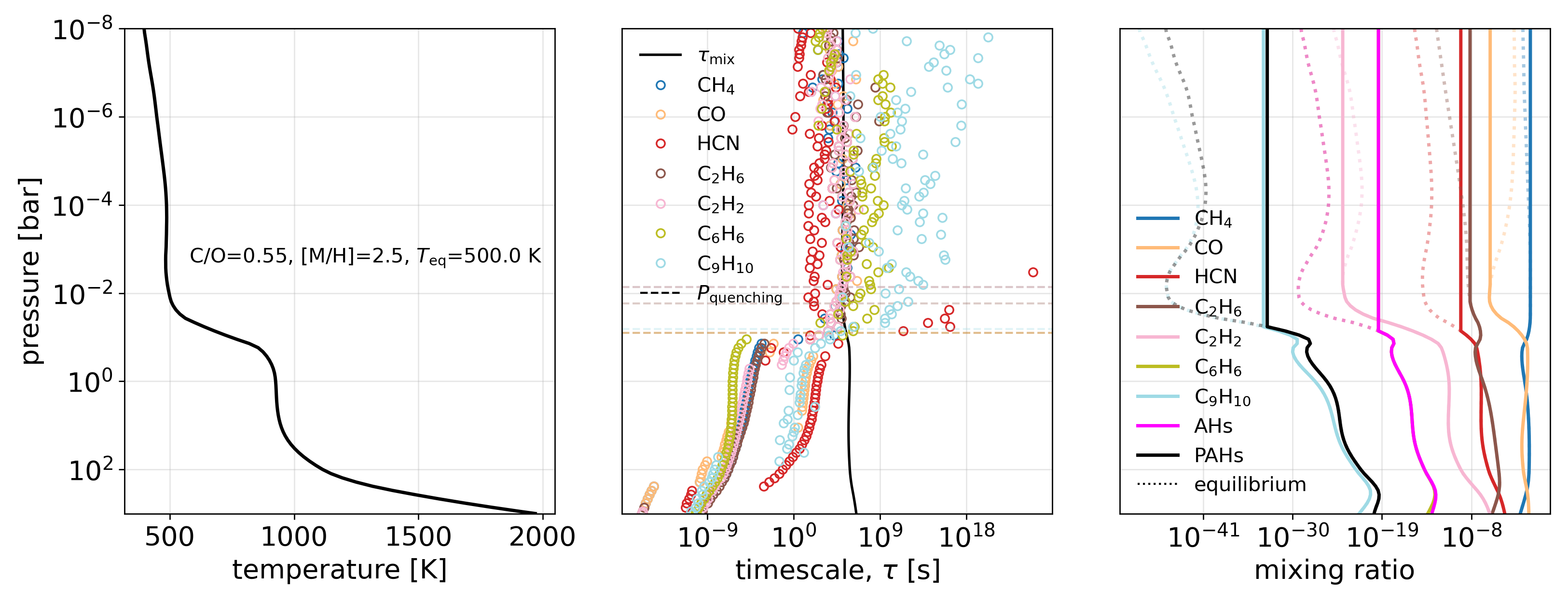}
     \caption{\footnotesize A representative model case with C/O=0.55, [M/H]=2.5, and $T_{\rm eq}=500$ K, shown as one example from the full grid of 360 simulations. (Left) $T$-$P$ profile computed using \texttt{HELIOS} \citep{malik2017helios}, as described in Section~\ref{subsec:helios}. (Middle) $\tau-P$ profiles computed using the characteristic chemical timescale approach described in Section~\ref{subsec:jacobian} and \ref{sec:appendix_evts}. The solid black line indicates the vertical mixing timescale, $\tau_{\rm mix}$; colored open symbols indicate the chemical timescales of individual chemical species; and colored dashed horizontal lines indicate the corresponding quench pressure points for each species. (Right) vertical mixing ratio profiles of chemical species. Colored solid lines show quenched abundances, while dotted lines show mixing ratios assuming thermochemical equilibrium. In this case, the chemical timescales of \ce{C6H6} (benzene) and \ce{C9H10} (indane) are used as the grid-representative chemical timescales for aromatic hydrocarbon (AH) and polycyclic aromatic hydrocarbon (PAH) species, respectively, as described in Section~\ref{subsec:classification}. Consequently, the AH (magenta solid line) and PAH (black solid line) group abundances inherit the quench pressure points of their corresponding grid-representative species. It should be noted that the vertical mixing ratio profiles of AHs and \ce{C6H6} nearly overlap; therefore, \ce{C6H6} is not shown in the figure.}
    \label{fig:showcase}
\end{figure*}

\section{Details of The RMG-Generated Chemical Network for Complex PAH Chemistry up to \ce{C16}}  \label{sec:rmg_appendix}

Although the Reaction Mechanism Generator (\texttt{RMG}) has been extensively validated in chemical engineering applications, its applicability to planetary atmospheres warrants further discussion. In general, chemical reaction networks are governed primarily by local temperature, pressure, and chemical composition, which determine reaction pathways and their corresponding rates. In this sense, the fundamental chemistry treated by \texttt{RMG} is applicable across a wide range of environments, including combustion systems, planetary atmospheres, and even the interstellar medium.

However, an important distinction is that planetary atmospheres are inherently multidimensional systems influenced by thermochemical processes in the deep atmosphere, vertical mixing in intermediate layers, and photochemistry in the upper atmosphere. The current implementation of \texttt{RMG} does not explicitly include photochemical processes or transport when generating reaction networks, which represents a key limitation of this approach. As a result, the network is best suited for describing thermochemical processes in the deep atmosphere rather than fully coupled photochemical–transport systems.

Despite this limitation, the present study focuses on deep atmospheric regions, where temperatures and pressures are high, and chemistry is largely governed by thermochemical kinetics and equilibrium. In these regions, quenching typically occurs at pressures where ultraviolet photons cannot penetrate \citep[e.g., $P\gtrsim17$ bar for Jupiter;][]{visscher2010deep}, and photochemical effects are therefore expected to be subdominant. Under these conditions, the \texttt{RMG}-generated networks provide an appropriate framework for modeling hydrocarbon growth and PAH formation in thermal chemistry-dominating deep atmospheres.

\texttt{RMG} has also been successfully applied in planetary and exoplanetary contexts. However, it is important to distinguish between validation of the chemical network construction and validation of the full atmospheric modeling framework. In several studies, such as WASP-39~b \citep{yang2024automated} and K2-18~b \citep{yang2024chemical}, \texttt{RMG}-generated networks were incorporated into more complete photochemical kinetic–transport models. These results demonstrate that \texttt{RMG} can generate chemically relevant reaction networks for planetary atmospheres, but do not directly validate the simplified framework adopted in this work.

The most direct validation of the specific thermochemistry + quench-timescale framework used here is the recent work application to Jupiter’s deep atmosphere, in which \texttt{RMG} was used to model CO$-$\ce{CH4} chemistry together with the same EVTS-based method to reproduce quenched CO constraints on the bulk oxygen abundance inferred from the latest Juno MWR analysis \citep{yang2026jupiter}. This agreement suggests that the EVTS approximation can capture the dominant quenching behavior in deep atmospheres, within the typical uncertainties of current atmospheric constraints.

We emphasize that the present framework remains an approximation to the full chemical kinetic–transport problem. In particular, it does not include fully coupled photochemistry or vertical transport, and therefore should not be interpreted as a fully predictive model of exoplanet atmospheres. Rather, it provides a computationally efficient and physically motivated approach for exploring trends in hydrocarbon growth and quenching in deep atmospheres.

To capture the complex chemistry of PAH formation, we benchmarked the most comprehensive CHO chemical network to date, spanning \ce{C2} to \ce{C16} \citep{liu2021predicting}. \citet{liu2021predicting} used the Reaction Mechanism Generator (\texttt{RMG}) \citep{Gao_2016, liu2021rmg, RMG-database} to automatically construct a detailed mechanism for acetylene (\ce{C2H2}) pyrolysis, predicting PAH formation from \ce{C2H2} to pyrene (\ce{C16H10}). This model has been shown to achieve good agreement with experimental data, with predicted mole fractions matching flow-reactor \ce{C2H2} pyrolysis experiments to within a factor of five for PAHs up to two rings (e.g., naphthalene and indene) at 8 kPa (80 mbar) over the temperature range 1073--1373 K \citep{norinaga-2008-c2h2}. The model also reproduced qualitative agreement to trends for three-ring PAHs (e.g., anthracene, phenanthrene, and acenaphthylene) and four-ring PAHs (e.g., pyrene), with predicted mole fractions agreeing within approximately four orders of magnitude, but did not capture PAHs larger than pyrene due to computational limitations in automatically generating larger PAH networks \citep{liu2021predicting}. These limitations include the rapidly increasing number of possible species and reactions, challenges in accurately treating highly strained intermediates, and uncertainties in thermochemical and kinetic data for higher-order growth pathways \citep{liu2021predicting}.

Because exoplanet atmospheres require chemistry initiated from \ce{C1} species, we did not directly use their original chemical network, which is initiated from \ce{C2} species \citep{liu2021predicting}. Instead, we adopted the \citet{liu2021predicting} \texttt{RMG} input largely unchanged, but added the \texttt{Klippenstein\_Glarborg2016} library to represent \ce{C1}--\ce{C2} chemistry, thereby generating a chemical network with RMG spanning \ce{C1} to \ce{C16} species. The \texttt{Klippenstein\_Glarborg2016} library is based on the methane-oxidation mechanism constructed by \citet{hashemi2016high}, which was extensively verified and benchmarked across a wide range of conditions using various experimental methane-oxidation studies \citep{hashemi2016high}. This library was previously incorporated to generate chemical networks for warm and hot Jupiter atmospheres \citep{yang2024automated}, for temperate sub-Neptune atmospheres \citep{yang2024chemical}, and for Jupiter's atmosphere \citep{yang2026jupiter}. In both applications, it successfully reproduced key atmospheric constraints for WASP-39~b, WASP-80~b, K2-18~b, and Jupiter \citep{yang2024automated, yang2024chemical, yang2026jupiter}. 
We assumed initial elemental abundances of C, H, and O equal to 1000$\times$ solar metallicity, temperatures of 300--2000 K, pressures of 1 mbar to 100 bar, and an integration time of 3.154$\times10^{19}$ s (10$^{12}$ yr), which is sufficient for the chemistry to reach steady state under these conditions. We generated the reaction network at 1000$\times$ solar metallicity because this high-metallicity composition yields a more comprehensive CHO network in \texttt{RMG}, retaining minor intermediates and reaction pathways that may not be included when the network is generated at lower metallicities due to flux-based inclusion criteria. The resulting network therefore remains applicable across lower-metallicity regimes (1–100$\times$ solar), where these pathways naturally become inactive if the corresponding species remain negligible. In principle, generating a separate reaction network for each metallicity in the model grid would be ideal; however, this approach is computationally prohibitive. Adopting a single, fixed reaction network ensures internal consistency across the full 1–10,000$\times$ solar metallicity grid while remaining computationally tractable. The \texttt{RMG} input file is provided in the Supplementary Materials (\texttt{input.py}). The resulting CHO network contains 656 species and 7027 reactions.

Exoplanet atmospheres also contain nitrogen and sulfur-bearing species that can potentially affect the characteristic chemical lifetimes of PAHs. It is therefore important to include these elements in the chemical network to accurately capture PAH chemistry and quenching behavior. To incorporate nitrogen- and sulfur-bearing chemistry, we merged this PAH network with the sub-Neptune chemical network of \citet{yang2024chemical} (92 species; 1666 reactions), removing duplicated species and reactions. As a result, the final merged network contains 700 species and 8258 reactions.


\section{Details of the Characteristic Chemical Timescale Approach}  \label{sec:appendix_evts}
We briefly summarize the EVTS (eigenvalue timescale method) implementation here, while full methodological details are provided in Section~2.7 and Appendix~F of \citet{yang2026jupiter}. The vertical mixing timescale is defined as
\begin{equation}
\tau_{\mathrm{mix}} = \frac{L^2}{K_{\mathrm{zz}}},
\end{equation}
where $L$ is the characteristic mixing length and $K_{\mathrm{zz}} = 10^6$ cm$^2$/s is adopted as a nominal eddy diffusion coefficient for temperate sub-Neptune atmosphere \citep{Zhang&Showman2018, yang2024chemical}. Here, $L$ is an effective mixing length scale associated with convective dynamics and should not be interpreted as the physical distance between atmospheric layers \citep{smith1998estimation}. We also tested two additional values of $K_{\mathrm{zz}}$ ($10^4$ and $10^8$ cm$^2$ s$^{-1}$), and found that the resulting trends are not strongly sensitive to the vertical mixing strength. Following mixing length theory, we assume $L = 0.1H$ for Neptune \citep{smith1998estimation}, where $H$ is the atmospheric scale height, computed assuming a gravitational acceleration of 10.65 m/s$^2$ adopted from that of GJ~1214~b \citep{Cloutier_2021}. It has to be noted that this mixing length theory is most appropriate for deep atmospheric regions of giant and sub-Neptune atmospheres, where the thermal structure is approximately convective or near-adiabatic.

In contrast, this approximation is not expected to be generally applicable in stably stratified radiative regions, such as terrestrial stratospheres, where the temperature gradient and transport regime differ substantially. In these cases, a full 1D photochemical kinetic--transport treatment is required to accurately capture the disequilibrium chemistry.

The gas temperature, T, and mean molecular weight, $\mu$ [g/mol] are required to compute the scale height $H$ and thus $\tau_{\mathrm{mix}}$. The former is taken as the local temperature in each atmospheric layer based on the precomputed $T$-$P$ profile, and the latter is calculated self-consistently from the thermochemical equilibrium composition at each atmospheric layer (detailed below). Chemical quenching is assumed to occur at the pressure level where $\tau_{\mathrm{chem}} = \tau_{\mathrm{mix}}$.

At each layer of the $T$–$P$ profile (117 layers in total; Section~\ref{subsec:helios}), we compute thermochemical equilibrium abundances using \texttt{Cantera} \citep{Goodwin_Cantera_An_Object-oriented_2024} with the chemical network described in Section~\ref{subsec:chemical_network}. Using the resulting thermochemical equilibrium species mole fraction profiles, we construct the Jacobian matrix, $J$, required for the EVTS calculation using \texttt{Cantera}'s \texttt{Kinetics.net\_production\_rates} \citep{Goodwin_Cantera_An_Object-oriented_2024}. In practice, this approach quantifies how small perturbations to the equilibrium chemical state relax back toward equilibrium through the coupled reaction network. By linearizing the chemical system around equilibrium, the Jacobian matrix captures how changes in the abundance of each species affect the net production of all other species. The eigenvalues of this Jacobian matrix, therefore, represent the characteristic decay timescale of independent chemical modes, from which we extract the dominant chemical timescale, $\tau_{\mathrm{chem}}(T,P)$, for each target species. 

The Jacobian matrix was constructed numerically using a central finite difference scheme, with each species' mole fraction perturbed by 50\% of its equilibrium value. From the resulting Jacobian eigenvalues, we compute $\tau_{\mathrm{chem}}(T,P)$ for the targeted species \ce{CO}, \ce{CH4}, \ce{C2H6}, \ce{C2H2}, \ce{HCN}, AHs (see Section~\ref{subsec:classification}), and PAHs (see Section~\ref{subsec:classification}).

\section{Planetary parameters used in this Work} \label{sec:parametric_data_table}
Table~D\ref{table:planet_data} lists the planetary parameters and corresponding references used in this work.

\begin{deluxetable*}{lcccl}[htb!]
\renewcommand{\tablename}{Table D}
\tablecaption{Planetary Parameters used in Figure~\ref{fig:2D-colormap-JWST}\label{table:planet_data}}
\tablehead{
\colhead{Planet (Spectrum Reference)} & \colhead{$T_{\rm eq}$ [K]} & \colhead{[M/H] [$Z_{\odot}$]} & \colhead{$A_H$[$H$\tablenotemark{a}]} & \colhead{Parameter References}}
\startdata
LP~791-18~c \citep[JWST;][]{roy2025diversity}& $355^{+2}_{-2}$ & $2.50^{+0.11}_{-0.11}$ & $0.02^{+0.05}_{-0.13}$ & \citet{peterson2023temperate, roy2025diversity}\\
GJ~9827~d \citep[JWST;][]{Piaulet-Ghorayeb_2024}& $675^{+14}_{-12}$ & $\geq$2.39 & $1.22^{+0.28}_{-0.36}$ & \cite{Piaulet-Ghorayeb_2024}\\
GJ~1214~b \citep[HST;][]{kreidberg2014clouds}& $567^{+8}_{-8}$ & $3.48^{+0.42}_{-0.36}$ & $0.09^{+0.03}_{-0.07}$ & \cite{Mahajan_2024,Ohno_2025}\\
GJ~3090~b \citep[JWST;][]{Ahrer_2025}& $693^{+18}_{-18}$ & $\geq$2.89 & $0.00^{+0.00}_{-0.03}$ & \cite{Almenara_2022_GJ3090b, Ahrer_2025}\\
GJ~3470~b \citep[HST;][]{benneke2019sub}& $615^{+16}_{-16}$ & $2.10^{+0.15}_{-0.15}$ & $1.03^{+0.26}_{-0.20}$ & \cite{Bonfils_2012_GJ3470b, Beatty_2024}\\
GJ~436~b \citep[HST;][]{knutson2014featureless}& $675^{+5}_{-5}$ & $\geq$2.70  & $0.58^{+0.32}_{-0.39}$ & \cite{mukherjee2025jwst}\\
TOI-836~c \citep[JWST;][]{wallack2024jwst}& $665^{+27}_{-27}$ & $\geq$2.24 & $-0.01^{+0.31}_{-0.13}$ & \cite{Hawthorn_2023_TOI-836c, wallack2024jwst}\\
TOI-421~b \citep[JWST;][]{davenport2025toi}& $922^{+14}_{-14}$ & $1.28^{+0.86}_{-1.15}$  & $3.75^{+0.45}_{-0.35}$ & \cite{krenn2024characterisation, davenport2025toi}\\
TOI-270~d \citep[JWST;][]{benneke2024jwst}& $387^{+10}_{-10}$ & $2.35^{+0.16}_{-0.24}$  & $2.87^{+0.34}_{-0.35}$ & \cite{van2021masses, benneke2024jwst}\\
TOI-674~b \citep[HST;][]{Brande_2022}& $635^{+15}_{-15}$ & $1.00^{+1.47}_{-1.00}$  & $1.05^{+0.46}_{-0.41}$ & \cite{murgas2021toi}\\
HD~3167~c \citep[HST;][]{Mikal-Evans_2021}& $511^{+18}_{-18}$ & $\geq$2.00  & $1.76^{+0.90}_{-0.89}$ & \cite{Howard_2025}\\
HD~106315~c \citep[HST;][]{Kreidberg_2022}& $799^{+14}_{-14}$ & $2.24^{+0.24}_{-0.24}$  & $1.91^{+1.14}_{-0.87}$ & \cite{Kreidberg_2022, Howard_2025}\\
HD~97658~b \citep[HST;][]{Guo_2020}& $751^{+12}_{-12}$ & $2.4^{+0.3}_{-0.4}$  & $0.06^{+0.74}_{-0.05}$ & \cite{Guo_2020, Ellis_2021}\\
HIP~41378~b \citep[HST;][]{brande2024clouds}& $959^{+9}_{-5}$ & $\geq$0.00 & $2.64^{+2.62}_{-2.50}$ & \cite{Howard_2025}\\
HAT-P-11~b \citep[HST;][]{fraine2014water}& $847^{+46}_{-54}$ & $\leq$1.93 & $2.69^{+0.59}_{-0.64}$ & \cite{Chachan_2019,basilicata2024gaps}\\
\enddata
\tablenotetext{a}{$H$ refers to the atmospheric scale height assuming a mean molecular weight of $\mu$=3.05 amu.}
\end{deluxetable*}


\bibliography{reference}{}
\bibliographystyle{aasjournal}



\end{document}